\documentstyle[12pt]{article}

\columnwidth\textwidth

\begin{document}

\newcommand{\reell}{{\kern+.25em\sf{R}\kern-.78em\sf{I}\kern+.78em\kern-.25em}}
\newcommand{\komplex}{{\sf{C}\kern-.46em\sf{I}\kern+.46em\kern-.25em}}
\newcommand{\posganz}{{\kern+.25em\sf{N}\kern-.86em\sf{I}\kern+.86em\kern-.25em}}
\newcommand{\ganz}{{\kern+.25em\sf{Z}\kern-.78em\sf{Z}\kern+.78em\kern-.65em}}
\newcommand{\Hamilton}{{\kern+.25em\sf{H}\kern-.86em\sf{I}\kern+.86em\kern-.25em}}
\newcommand{\Cayley}{{\sf{O}\kern-.56em\sf{I}\kern+.56em\kern-.25em}}
\newcommand{\unit}{{\sf{1}\kern-.18em\sf{I}\kern+.18em\kern-.18em}}
\newcommand{\opunit}{{\sf{1}\kern-.29em\sf{1}\kern+.29em\kern-.33em}}
\newcounter{blabla}
\newcommand{\be}{\begin{equation}}
\newcommand{\ee}{\end{equation}}
\newcommand{\bdm}{\begin{displaymath}}
\newcommand{\edm}{\end{displaymath}}
\newcommand{\beann}{\begin{eqnarray*}}
\newcommand{\eeann}{\end{eqnarray*}}
\newcommand{\bea}{\begin{eqnarray}}
\newcommand{\eea}{\end{eqnarray}}

\newcommand{\nn}{\nonumber \\}

\newcommand{\resection}[1]{\setcounter{equation}{0}\section{#1}}
\renewcommand{\theequation}{\arabic{equation}}
\newcommand{\appsection}{\addtocounter{section}{1}
           \setcounter{equation}{0}\section*{Appendix\Alph{section}}}

\begin{titlepage}

\def\mytoday#1{{ } \ifcase\month \or
 January\or February\or March\or April\or May\or June\or
 July\or August\or September\or October\or November\or December\fi
%\space\number\day ,
 \space \number\year}
\noindent
\hspace*{11cm} BUTP--94/21\\
\vspace*{0.5cm}
\begin{center}
{\LARGE Group Quantization of Parametrized Systems. I. Time Levels}

\vspace{1cm}

P. H\'{a}j\'{\i}\v{c}ek
\\
Institute for Theoretical Physics \\
University of Bern \\
Sidlerstrasse 5, CH-3012 Bern, Switzerland
\\ \vspace{1cm}

March 1995 \\ \vspace*{1cm}

\nopagebreak[4]

\begin{abstract}
A method of quantizing parametrized systems is developed that is
based on a kind of ``gauge invariant'' quantities---the so-called
perennials (a perennial must also be an ``integral of motion'').
The problem of time in its particular form (frozen time formalism,
global problem of time, multiple choice problem) is met, as well as
related difficulty characteristic for this type of theory: the
paucity of perennials. The present paper is an attempt to find some
remedy in the
ideas on ``forms of relativistic dynamics'' by Dirac. Some aspects
of Dirac's theory are generalized to all
finite-dimensional first-class parametrized systems. The
generalization is based on replacing the
Poincar\'{e} group and the algebra of its generators as used by Dirac
by a
canonical group of symmetries and by an algebra of elementary
perennials. A number of insights is gained; the following are the
main results. First, conditions are revealed under which the time
evolution of the ordinary quantum mechanics, or a generalization of
it, can be constructed. The construction uses a kind of gauge and
time choice and it is described in detail.
Second, the theory is structured so that the quantum mechanics
resulting from different choices of gauge and time are compatible.
Third, a practical way is presented of how a broad class of problems
can be solved without the knowledge of explicit
form of perennials.

\end{abstract}

\end{center}

\end{titlepage}

\section{Introduction and summary}

         Systems with constraints are frequently met in the
         contemporary theoretical physics. One can distinguish between
         two quite different cases of these systems: the so-called
         {\em gauge systems} whose classical solutions (curves in the
         phase space) are transversal to the orbits of the gauge group
         (which is generated via Poisson brackets by the constraints),
         and the so-called {\em parametrized systems} each solution
         curve of which lies within an orbit. We will concentrate on
         some typical problems associated with the parametrized
         systems.

         The most challenging parametrized system is Einstein's
         general theory of relativity. Attempts to quantize this
         theory meet a number of difficulties of a technical as well
         as conceptual character (for reviews, see [1], [2]
         and[3]). Many conceptual problems are more or less
         directly associated with the notion of time -- which is an
         issue common to all parametrized systems. On one hand, the
         choice of time is to a certain degree analogous to the choice
         of gauge in a gauge theory; on the other hand, the time plays
         a very special role in quantum mechanics and it surely
         has some measurable aspects. This dichotomy is the source of
         several problems.

         In particular, two typical problems can arise: the so-called
          ``global time problem'' and the ``multiple choice problem''
         (for details and examples, see [1] and [2]). The
         essence of the former, which is reminiscent of the Gribov
         problem, is that given constraints need not admit a global
         choice of time. In the case that there is a global time, we
         meet the latter problem -- there will be no uniqueness in the
         choice of time and different choices together with the
         quantization method at hand will often lead to unitarily
         inequivalent quantum theories.

         The multiple choice problem may be related to another,
         much more general problem: a given classical theory does not
         determine ``the corresponding quantum theory'' uniquely.
         There are (at least) factor ordering ambiguities. However,
         the ``spirit of symmetry'' of the classical theory suggests
         that we are to look for such a factor ordering that the
         resulting quantum theory becomes as invariant as the original
         classical theory. A pioneer effort in this direction has been
         undertaken in ref. [4].

         The present paper will describe another attempt of this kind;
         it will be based on an old paper [5] by Dirac. In fact,
         [5] was aimed at a special relativistic {\em mechanics}
         of elementary particles that avoids the infinities of the
         quantum field theory. Of course, the sweeping success of the
         quantum field theory as we know it today pushed the Dirac
         paper into oblivion (but see [6]). However, it turns out
         that some ideas of ref. [5] have a direct relevance to
         quantization of parametrized systems. They lead to clean
         separation between those structures of the theory that are
         independent of possible time choices (and which are
         considered more fundamental) and many possible time
         structures which can be constructed after the fundamental
         structure is given.

         Dirac's fundamental structure was some sufficiently complete
         algebra of, or group generated by, observables. Observables
         were defined as those phase space functions that have weakly
         vanishing Poisson brackets with the constraints. For the
         special relativistic system considered by Dirac, these
         observables were naturally associated with the Poincar\'{e}
         group. Today, analogous algebras and groups are used for much
         broader class of parametrized systems: in their work on
         algebraic quantization, Ashtekar and his collaborators (see
         [7] and [8]) employ also this kind of algebras; the
         so-called group quantization ([9], [10] and [11]) can be
         based on similar groups [12]. In fact, both methods are
         related [8], and they could be combined.

         Thus, Dirac's theory [5], if obviously generalized, will
         include some aspects of algebraic and group quantization; we
         will see that it includes both quantizations in a certain
         sense. However, it is much more to it! Indeed, the next
         important Dirac's idea is the so-called ``time surface''
         in the phase space. Exactly as a gauge surface of any gauge
         system, the time surface has to be transversal to the orbits
         generated by the constraints. The observables and the group
         action are projected onto a given time surface. The resulting
         functions on, or transformations of, the time surface have
         the same algebraic, or group, structure independently of how
         the time surface is chosen. In such a manner the desired
         time-choice independence is achieved.

         The most interesting idea by Dirac concerns the time
         evolution. Any element of the algebra of observables which,
         via Poisson brackets, moves the time surface non-trivially in
         the phase space is eligible as a ``Hamiltonian''. Thus one
         obtains a family of ``time levels'' in the phase space such
         that each two time levels can be mapped onto each other by a
         symmetry. Using this symmetry, we can easily give meaning to
         ``the same measurement at different times''.
         The only role of the Hamiltonian is to
         generate (to define) these structures; it need not be the
         total energy of the system, it need not be positive, or
         define a ground state. Dirac, as well as subsequent work like
         e.g. [6] showed that the idea works for well-known special
         relativistic systems. This generalization
         of time evolution can be felt as as liberating by those
         who try to quantize gravity.

         There is a double freedom in the construction of time
         evolution \`{a} la Dirac: The choice of a time surface and that
         of an associated Hamiltonian. Formally, quantum mechanics
         constructed for different choices of Hamiltonian will not be
         unitarily equivalent. This can be shown even for the free
         relativistic particle [13]. This is a ``remnant'' of the
         multiple choice problem in Dirac's theory. However, it seems
         that this feature can be controlled and does not lead to any
         serious problem. The theories are ``compatible'', if not
         unitarily equivalent [13].

         In the present paper, we will generalize Dirac's theory to
         all finite-dimensional first-class parametrized systems. The
         restriction to finite number of degrees of freedom
         is chosen just for the sake of
         simplicity and because it enables a relatively rigorous
         exposition. The plan of the paper is as follows.

         In Sec. 2, we describe the geometrical properties of the
         constraint surface of the first-class systems following
         essentially ref. [14]. Then, we give a definition of the
         parametrized system that is in agreement with this
         geometrical point of view and describe in some detail the
         meaning of curves within orbits at the constraint surface, as
         well as the relation of orbits to certain maximal classical
         solutions.

         Sec. 3 defines the basic notions of observables and
         symmetries of a parametrized system. In fact, the name
         ``observables'' for the variables that are constant along
         orbits is misleading in the case of parametrized systems
         [15]. There are observable features of these systems that
         cannot be described by these observables; in particular,
         Dirac's time structures are clearly such features. Thus, we
         will adhere to the proposal of [15] using the word
         ``perennial'' instead of ``observable''. A perennial is
         defined as
         a function that is constant along orbits and a symmetry is a
         symplectic diffeomorphism that preserves the constraint
         surface (see also [12]). We study relations between
         perennials and symmetries, show some general properties and
         define the so-called ``algebras of elementary perennials''
         and ``first-class canonical groups'' for parametrized
         systems (differing by some detail from [8] or [12]).
         By the way, extended
         use of perennials in quantum gravity has been criticized in
         [15], because no single perennial is known for general
         relativity, and because the general relativity has no
         perennials of a certain kind [16]. However, an abstract
         existence of general perennials for any parametrized system
         is at least very plausible (see, e.g. [17] and an argument in
         Sec. 4 of the present paper). Such an existence will be
         often sufficient for our aims: as we shall see, Dirac's
         method avoids the necessity to know perennials explicitly
         for many important calculations.

         In Sec. 4, the so-called ``transversal surfaces'' are defined
         (we replace Dirac's name ``time surface'' by it). Perennials
         and symmetries are projected to transversal surfaces and the
         projection is studied. We show that the Poisson algebra
         structure as well as the group structure is preserved by the
         projection. The projection also preserves the relation
         between one-dimensional subgroups of symmetries and the
         corresponding perennial generators. Thus, if the projections
         of perennials
         and symmetries are known (which is a common situation),
         practical calculations can be performed without knowledge
         of the functional form of the perennials on the whole phase
         space.

         In Sec. 5, we describe an obvious generalization of Dirac's
         notion of Hamiltonian to any parametrized system. The crucial
         role is played by curves within the orbits and their
         relation to classical solutions. Using this relation we can
         define Hamiltonians or partial Hamiltonians associated with a
         given transversal surface. Any partial Hamiltonian generates
         a one-dimensional group of symmetries that moves the
         corresponding transversal surface in the phase space. In this
         way, we obtain time levels in the phase space and time slices
         in the classical solutions of the system. The time slices
         generated by partial Hamiltonian cover only some open part of
         some classical solution. A complete family of partial
         Hamiltonians generate a more-dimensional family of time
         slices that cover all classical solutions completely. A
         subfamily of the full system of functional-time slices
         results in this way. Any two time levels in the phase space
         that have been generated from a given transversal surfaces
         by a
         complete family of partial Hamiltonians are related to each
         other by a symmetry transformation. This transformation can
         be made to map perennials into perennials and thus define
         related measurements at different time levels. This is used
         to develop a quantitative and complete theory of changes in
         time. The corresponding classical dynamical equations are
         derived. The way to the Schr\"{o}dinger and Heisenberg
         picture of quantum dynamics is straightforward. The families
         of perennials generated by partial Hamiltonians define some
         particular class of ``evolving constants of motion'' ([18]).
         Again, the time evolution of any variable can be calculated
         from the projection of the relevant perennials and the
         Hamiltonians to a transversal surface and so practical
         calculations do not need the complete knowledge of the
         perennials.

         The Appendix A illustrates the theory by working out
         explicitly some of our new notions and theorems for a toy model: the
         system of free massive particles in Minkowski space. Finally,
         proofs of some theorems are collected in the Appendix B.

         \section{First-class parametrized systems}

         In this section we give basic definitions, explain the
         notation and recall some well-known facts about
         finite-dimensional first-class parametrized systems. Most
         examples of such systems that can be met in the literature
         are usually presented in the following form.

         An $N$-dimensional manifold ${\cal C}$ is specified as the
         {\em
         configuration space}, and its cotangent bundle $T^\star {\cal
         C} = \Gamma$ as the {\em phase space} of the system. Some
         coordinates $\{q^\mu\}$ are chosen on $C$, and the
         corresponding coordinates $\{q^\mu, p_\mu\}$ on $\Gamma$.
         Then, the action $S$ of the system is written as follows:
         $$
         S = \int d\tau (p_\mu \dot{q}^\mu - N_\alpha  C_\alpha), \eqno(1)
         $$
         where the dot means the derivative with respect to $\tau,
         N_\alpha$
         are {\em Lagrange multipliers} (additional independent
         variables) and $ C_\alpha, \alpha = 1, \ldots, \nu'$,
         are the {\em constraint functions} (shortly ``constraints'').

         Varying $S$ with respect to $N_\alpha$'s, we obtain the
         following system of eqs.:
         $$
         C_\alpha = 0 \hspace{0.5cm} \forall_\alpha, \eqno(2)
         $$
         the so-called {\em constraints}. The system (2) defines a
         subset $\tilde{\Gamma}$ of $\Gamma$ which is called {\em
         constraint surface}.

         The system is called {\em first-class}, if the constraints
         satisfy the following conditions
         $$
         \{C_\alpha, C_\beta\} = f_{\alpha \beta \gamma} C_\gamma
         \hspace{5mm} \forall \alpha, \beta, \eqno(3)
         $$
         where $f_{\alpha \beta \gamma} (q, p)$ are some regular
         functions on $\Gamma$ and the symbol $\{ .,. \}$
         denotes the Poisson bracket. It follows that
         $$
         \{C_\alpha, C_\beta\}\mid_{\tilde{\Gamma}} = 0 \hspace{5mm}
         \forall \alpha, \beta; \eqno(4)
         $$
         one can show (see, e.g. [19]) that the condition (4) implies
         eqs. (3).

         \subsection{Geometry of constraint surface}

         We will need a geometrical definition of first-class
         systems, as it is given e.g. in [14]

         In this approach, the basic space is the phase space
         $\Gamma$. $\Gamma$ need not be a cotangent bundle; one just
         assumes $\Gamma$ to be a symplectic manifold. That is,
         $\Gamma$ is a $2N$-dimensional manifold equipped with a
         non-degenerate closed two-form $\Omega$. For the system
         defined by the action (1), $\Omega = dp_\mu \wedge dq^\mu$. In
         local coordinates $z^A$ on $\Gamma$, $A = 1, \ldots, 2N$,
         $\Omega$ can be represented by an anti-symmetric tensor field
         $\Omega_{AB}$ with the properties
         \begin{enumerate}
         \item $\Omega_{AB}(x) = -\Omega_{BA}(x), \hspace{5mm} \forall
         x \in \Gamma, A, B$
         \item $\Omega_{AB}(x)$ has an inverse matrix, $\Omega^{AB}(x)$,
         at each $x \in \Gamma$,
         \item $\Omega_{AB,C} + \Omega_{BC,A} + \Omega_{CA,B} = 0$ on
         $\Gamma$,
         \item Poisson bracket of two functions $f$ and $g$ on
         $\Gamma$ is defined as follows
         $$
         \{f,g,\} = \Omega^{AB} f_{,A} g_{,B}.
         $$
         \end{enumerate}
         The basic geometrical object of this approach is the
         constraint surface $\tilde{\Gamma}$. If $\tilde{\Gamma}$ is
         defined by eqs. (2), then, for many interesting systems,
         $\tilde{\Gamma}$ will not be a surface: the rank of the matrix

         \setcounter{equation}{4}
         \be
         \left ( \begin{array}{lll}
         C_{1,1},& \ldots, &C_{1,2N} \\
         \vdots &          & \vdots \\
         C_{\nu',1}, &\ldots, &C_{\nu',2N}
         \end{array}
         \right)
         \ee
         will not be constant along $\tilde{\Gamma}$, in some open
         dense subset of $\tilde{\Gamma}$, it will be $\nu$, say, and
         in the rest of points, the so-called critical points, it will
         be lower than $\nu$. In the critical points, $\tilde{\Gamma}$
         will have cusps and singularities.

         However, if one ``cuts out'' the critical points, then a
         smooth surface of dimension $2N-\nu$ results. Such a surface
         will, of course, not be closed in $\Gamma$. We will define
         the constraint surface $\tilde{\Gamma}$ in this way, and we
         will assume that a reasonable quantum theory can still be
         constructed (some discussion of this problem is given in
         [20]).

         Finally, we have to express the conditions (3) or (4) by some
         geometrical property of $\tilde{\Gamma}$. This can be done as
         follows (see [14]).

         Let $p \in \tilde{\Gamma}$. We define $T_p \tilde{\Gamma}$ as
         the set of all vectors from $T_p \Gamma$ that are tangential
         to $\tilde{\Gamma}$. $T_p \tilde{\Gamma}$ is a
         $(2N-\nu)$-dimensional subspace of $T_p \Gamma$. Then, the
         linear space $N_p$ at each $p \in \tilde{\Gamma}$ can be
         defined as follows:
         $$
         N_p = \{n \in T^\star_p \Gamma \mid < n, X> = 0, \hspace{2mm}
         X \in T_p \tilde{\Gamma} \}.
         $$
         $N_p$ is a $\nu$-dimensional subspace of $T^\star \Gamma$, and it
         is spanned by the gradients (5) of the constraint functions
         (for the system defined by the action (1)). Finally, define
         the linear space $X_p$ at all points of $\tilde{\Gamma}$ by
         $$
         X_p = \{ X \in T_p \Gamma \mid X^A = \Omega^{AB}n_B, n_B \in
         N_p \}.
         $$
         As $\Omega$ is non-degenerate, $X_p$ is a $\nu$-dimensional
         subspace of $T_p \Gamma$. $X_p$ is called {\em longitudinal
         subspace} of $T_p\tilde{\Gamma}$, and its elements are {\em
         longitudinal vectors}.

         With these definitions, we have the following theorem:

         {\bf Theorem 1}: $\tilde{\Gamma}$ is a constraint surface of
         a first-class system, iff
         $$
         X_p \subset T_p \tilde{\Gamma}. \eqno(6)
         $$
         Then, $X_p$ is an integrable distribution in
         $\tilde{\Gamma}$.

         A proof of the Theorem 1 is given in [14]. The condition (6)
         can be used as a geometrical definition of the first-class
         systems. The longitudinal vectors have the following property.
         Let $\tilde{\Omega}$ be the pull-back of $\Omega$ to
         $\tilde{\Gamma}$. Then,
         $$
         X_p = \{X \in T_p \tilde{\Gamma} \mid \Omega (X,Y) = 0,
         \hspace{2mm} \forall \hspace{2mm} Y \in T_p\tilde{\Gamma}\}.
         \eqno(7) $$
         Thus, $X_p$ is the subspace of degeneracy of
         $\tilde{\Omega}$. Indeed,
         $$
         \Omega_{AB} X^A Y^B = \Omega_{AB} \Omega^{AC} n_C Y^B = n_C
         Y^C = 0.
         $$
         As the distribution $X_p$ is integrable, there will be an
         integral manifold to $X_p$ through any point $p$ of
         $\tilde{\Gamma}$. Let us call the maximal integral manifolds
         of $X_p$ in $\tilde{\Gamma}$ {\em orbits}. Orbits are
         $\nu$-dimensional submanifolds of $\tilde{\Gamma}$ such that
         each point $p \in \tilde{\Gamma}$ lies at exactly one orbit,
         $\gamma_p$. The quotient, $\tilde{\Gamma}/\gamma$, is the
         physical phase space; it has dimension $2N-2\nu$.

         \subsection{Classical solutions}

         Here, we describe the relation between classical solutions
         and orbits.

         In a pure gauge theory, the constraints generate, via Poisson
         brackets, the gauge transformations. Any pair of points of a
         given orbit can be joined by a transformation which is
         generated by the constraints. Hence, an orbit is a collection
         of all gauge equivalent initial data of the gauge theory. For
         parametrized systems, the situation is different. Indeed,
         varying the action (1) with respect to $q^\mu$ and $p_\mu$, we obtain
         the following eqs.:
         $$
         \dot{q}^\mu = N_\alpha \frac{\partial C_\alpha}{\partial
         p_\mu} = N_\alpha \{q^\mu, C_\alpha\}, \eqno(8a)
         $$
         $$
         \dot{p}_\mu = - N_\alpha \frac{\partial
         C_\alpha}{\partial
         q^\mu} = N_\alpha \{ p_\mu, C_\alpha\}. \eqno(8b)
         $$
         Hence, the tangential vector $(\dot{q}^\mu, \dot{p}_\mu)$ to
         any classical solution $(q^\mu(\tau), p_\mu(\tau))$ lies in
         $X_{(q,p)}$ (as it must lie at $\tilde{\Gamma}$). Thus,
         classical solutions lie inside orbits.

         In general, a classical solution consits of a {\em base
         manifold} ${\cal M}$ of a fixed dimension, and a system $Q$
         of fields on ${\cal M}$. Two classical solutions are
         considered as equal, if their base manifolds are diffeomorph
         and if the corresponding fields can be brought into
         coincidence by means of this diffeo plus a gauge
         transformation.

         Given a parametrized system, then there is also an information
         on how a classical solution $({\cal M}_\sigma, Q_\sigma)$ is
         to be constructed for a curve $\sigma: \reell \rightarrow
         \gamma$ in an orbit $\gamma$. This information is different
         from system to system, but we will assume that it has the
         following general properties:

         Each orbit $\gamma$ defines a unique maximal solution $({\cal
         M}_\gamma, Q_\gamma)$ such that
         \begin{enumerate}
         \item there is a curve $\sigma: \reell \rightarrow \gamma$
         satisfying $({\cal M}_\sigma, Q_\sigma) = ({\cal M}_\gamma,
         Q_\gamma)$;
         \item if $\sigma'$ is an arbitrary curve in $\gamma$, then
         the corresponding classical solution $({\cal M}_{\sigma},
         Q_{\sigma})$ is a
         part of $({\cal M}_\gamma, Q_\gamma): {\cal M}_{\sigma}
         \subset {\cal
         M}_\gamma, \hspace{2mm}Q_{\sigma} = Q_\gamma
         \mid_{{\cal M}_{\sigma}}$;
         \item the classical solutions $({\cal M}_\gamma, Q_\gamma)$
         and $({\cal M}_{\gamma'}, Q_{\gamma'})$ defined by two
         different orbits $\gamma$ and $\gamma'$ are different.
         \end{enumerate}
         An arbitrary curve $\sigma'$ in $\gamma$ need not correspond
         to any non-trivial piece $({\cal M}_\gamma, Q_\gamma)$: for
         example, a constant curve $\sigma(\tau) = p \in \gamma,
         \hspace{2mm}
         \forall \hspace{2mm} \tau \in \reell$ represents just
         initial data
         for $({\cal M}_\gamma, Q_\gamma)$. We will say that a curve
         $\sigma'$ in $\gamma$ represents a non-trivial solution, if
         ${\cal M}_{\sigma'} \subset {\cal M}_\gamma$ contains an open
         subset of ${\cal M}_\gamma$.

         In general relativity (which is not an example of our
         parametrized systems as it is infinite-dimensional, but the
         situation is analogous), the base manifold of the classical
         solutions is a fixed four-dimensional manifold ${\cal M}$,
         and the system of fields is the four-metric $g_{\mu \nu}(x)$
         on ${\cal M}$. $g_{\mu \nu}(x)$ is constructed from the curve
         $^3g_{kl}(x,t), \pi^{kl}(x,t)$ in
         an orbit $\gamma$ by calculating first the corresponding laps
         and shift multipliers $N$ and $N^k$ from equations analogous to
         (8), and then constructing $g_{\mu \nu}(x)$ from
         $^3g_{kl}(x), N(x)$ and $N^k(x)$ by the
         well-known formula (see, e.g. [21], p. 507). The maximal
         solution $({\cal M}_\gamma, Q_\gamma)$ that is associated
         with an orbit $\gamma$ will be a maximal dynamical evolution
         of initial data; thus ``maximality'' is one in the sense
         of [22] rather than inextendibility (see, e.g. [23]). Then,
         two maximal solutions corresponding to two different orbits
         can overlap. Examples are obtained by studying spacetime
         solutions which are not globally hyperbolic.

         In Appendix A, the system of $\nu$ free massive particles in
         Minkowski space is considered. The relation between orbits
         and classical solutions is easily constructed and the
         properties 1)- 3) are proved.

         \section{Perennials and symmetries}

         \subsection{Algebras of elementary perennials}

         In ref. [7] and [8], the so-called {\em algebraic
         quantization} method is described. The emphasis is on the
         practical methodical aspects; this gives the ``quantization''
         a form of a procedure split in quite a number of intermediate
         steps. Our aim is different: we are going to study some
         general properties of the quantum theory constructed in this
         way and to perform some additional constructions. For
         example, we are interested in relations between the Ashtekar
         and Dirac approaches, or the Ashtekar and group approaches
         (see the next subsection), and in constructing time evolution
         and
         Hamiltonians. For this purpose it is advantageous to suppress
         a number of the procedural steps so that only some relevant
         properties of the end result of the procedure remain.

         The basic notion of our approach to algebraic quantization
         will be that of {\em perennial}. What is a perennial?

         In ref. [24], Dirac introduced the notion of the
         {\em first-class quantity} for constraint systems. In the
         current literature, the name {\em observable} is generally
         used for these quantities. However, in the case of
         parametrized systems, there seem to be observable aspects
         which are not described by first-class quantities; the
         first-class quantities form a rather narrower class of
         integrals of motions. These observations lead in [15] to
         introduce the name {\em perennials} for the first-class
         quantities of parametrized systems. This enables one to
         distinguish between perennials and observables; we will find
         such a distinction very useful, and we will adhere to
         Kuchar's nomenclature.

         For a system (1), a perennial $o $ is a (complex)
         $C^\infty$-function on the phase space such that
         $$
         \{o, C_\alpha\} \mid_{\tilde{\Gamma}} = 0, \hspace{5mm}
         \forall \alpha.
         $$
         It follows that $o$ is constant along each orbit and, vice
         versa any $C^\infty$-function that is constant along orbit is
         a perennial. We can use this property as a more geometrical
         definition of perennials.

         Let $o_1$ and $o_2$ be two perennials. Then, their complex
         linear combination $\lambda_1 o_1 + \lambda_2 o_2$, their
         product $o_1o_2$, their Poisson bracket $\{o_1,o_2\}$, and
         their complex conjugation $\bar{o}_1$ and $\bar{o}_2$ are
         again perennials. Any set of perennials that is closed with
         respect
         to these four operations is called {\em Poisson algebra (with
         involution)}.

         One can observe that any constraint function $C_\alpha$
         satisfies the definition of perennial. However, from the
         physical point of view, these perennials are trivial being
         equal to zero on $\tilde{\Gamma}$. Still, the Poisson algebra
         of all perennials ${\cal P}$ will contain all constraint
         functions together with the whole ideal ${\cal J}_c$ which is
         generated by them. Thus, the quotient space
         $$
         \tilde{{\cal P}} = {\cal P}/{\cal J}_c
         $$
         will again form a well-defined Poisson algebra (with
         involution, if $\bar{{\cal J}}_c = {\cal J}_c$). Any two
         perennials in the same class of $\tilde{{\cal P}}$ are equal
         on $\tilde{\Gamma}$, so they are not distinguishable from the
         physical point of view.

         Consider a subset $\tilde{{\cal S}}$ of $\tilde{{\cal P}}$
         with the following properties.
         \begin{description}
         \item (i) $\tilde{{\cal S}}$ is a Lie algebra with involution,
         that is, $\tilde{{\cal S}}$ is closed with respect to complex
         linear
         combination, Poisson brackets, and complex conjugation.

         \item (ii) $\tilde{{\cal S}}$ has so many elements that they
         almost
         separate the
         orbits. That is, if $\gamma_1$ and $\gamma_2$ are two
         orbits,
         such that there is a perennial $o \in {\cal P}$ satisfying
         $$
         o (\gamma_1) \neq o(\gamma_2),
         $$
         then there is a class of perennials $\{o'\} \in \tilde{{\cal
         S}}$ such that
         $$
         o'(\gamma_1) \neq o'(\gamma_2).
         $$
         (For many parametrized systems, there will be pairs of
         different orbits which cannot be separated by any continuous
         function of orbits; that is, $\tilde{\Gamma}/ \gamma$ is
         non-Hausdorff. For examples see e.g. [25]).
         \end{description}
         $\tilde{{\cal S}}$ will be called {\em algebra of elementary
         perennials}.
         In general, $\tilde{{\cal S}}$ is not uniquely determined.
         The construction of the quantum mechanics will, however,
         depend on the choice of $\tilde{{\cal S}}$. Thus, the choice
         of $\tilde{{\cal S}}$ reflects the non-uniqueness of
         construction of quantum mechanics from a classical theory.
         This non-uniqueness has, however, still another aspect.
         $\tilde{{\cal S}}$ can contain Lie subalgebras with
         involution which still almost separate orbits. These can be
         called {\em subalgebras of elementary perennials}.

         A given Lie algebra $\tilde{{\cal S}}$ can contain elements
         that are not independent algebraically. For example, there
         can be $f,g,h$ in $\tilde{{\cal S}}$ satisfying a {\em
         relation} in $\tilde{{\cal P}}$
         $$
               h =fg. \eqno(9)
         $$
         For topologically nontrivial $\tilde{\Gamma}/\gamma$, there
         always will be such relations.

         A {\em quantization} can now be defined as any faithful
         representation of the Lie algebra $\tilde{{\cal S}}$ in a
         Hilbert space ${\cal H}$ satisfying the following
         requirements. Let us denote by $\hat{f}$ the linear operator
         on ${\cal H}$ that represents the element $f \in \tilde{{\cal
         S}}$. Then
         \begin{description}
         \item (a) $\widehat{(\lambda f + \kappa g)} = \lambda \hat{f} +
         \kappa
         \hat{g}$,
         \item (b) $\widehat{\{f,g\}} = - i [\hat{f}, \hat{g}]$,
         \item (c) $\hat{\bar{f}} = \hat{f}^\dagger$,
         \item (d) if a constant function ``1'' with value 1 is in
         $\tilde{{\cal S}}$, then
         $$
         \hat{1} = id,
         $$

         \item (e) all relations are satisfied, if the products are
         replaced by symmetrized products; for example, if $(9)$
         holds, then
         $$
         \hat{h} = \frac{1}{2} \hat{f} \hat{g} + \frac{1}{2} \hat{g}
         \hat{f}, \eqno(10)
         $$

         \item (f) all subalgebras of elementary perennials in
         $\tilde{{\cal S}}$ are {\em irreducibly} represented in
         ${\cal H}$.
         \end{description}

         Again, there will be more such representations (unitarily
         inequivalent). Thus, the choice of the representation is
         another freedom in the construction of quantum mechanics.

         Observe that the requirement of irreducibility concerns all
         subalgebras of elementary perennials in $\tilde{{\cal S}}$.
         This implies, of course, that $\tilde{{\cal S}}$ also is
         represented irreducibly. This requirement is analogous to the
         usual requirement for non-constrained systems: any physical
         representation of some Lie algebra of phase space functions
         must induce an irreducible representation of the Heisenberg
         algebra (see, e.g. the discussion of the Van Hove theorem in
         [26], p.435).

         \subsection{Canonical groups}

         A different method of quantization, which is however, quite
         closely related to the algebraic one, is the {\em group
         quantization method}. In refs. [9], [10] and [11], the method
         is explained for non-constrained systems, or only few first
         procedural steps are given for the constrained ones. In ref.
         [12], a sort of short-cut of the group method is given for
         constrained systems that is somewhat analogous to our
         previous subsection. In the present subsection we will
         describe a minor modification of Rovelli's approach [12] that
         will be suitable for our purposes.

         The basic notion of group quantization is that of a {\em
         symmetry} of a parametrized system.

         {\bf Definition}: Let $\varphi : \Gamma \rightarrow \Gamma$
         satisfy
         the conditions:
         \begin{enumerate}
         \item $\varphi$ is a diffeo with a domain $\mbox{D}
         (\varphi)$ dense in $\Gamma$,
         \item $\varphi$ is a symplectic map, $\varphi^\star \Omega =
         \Omega$,
         \item $\varphi$ preserves the constraint surface, $\varphi
         \tilde{\Gamma} \subset \tilde{\Gamma}$;
         \end{enumerate}
         then $\varphi$ is called a {\em symmetry} of the parametrized
         system.

         Hence, the symmetries of a parametrized system just have to
         preserve the constraint surface -- those of a gauge system
         have to preserve the Hamiltonian in addition.

         Observe also that we admit symmetries that can become
         singular somewhere in $\Gamma$ (and even on
         $\tilde{\Gamma}$).

         The following theorem will often be used

         {\bf Theorem 2}:
         \begin{enumerate}
         \item Let $\varphi$ be a symmetry. Then $\varphi$ maps orbits
         onto orbits.
         \item If the vector field $\xi$ on $\Gamma$ is an
         infinitesimal symmetry, then for any $p \in \tilde{\Gamma}
         \cap \mbox{Dom}\xi$ it holds:

         \begin{description}
         \item a) $\xi (p) \in T_p \tilde{\Gamma}$

         \item b) there is a neighbourhood $U$ of $p$ in $\Gamma$ and
         a function $F:U \rightarrow \komplex$ such that $\xi^A =
         \Omega^{AB} \partial_B F$ in $U$;

         $F$ satisfies the equation
         $$
         \{F, C_\alpha\}\mid_{\tilde{\Gamma}} = 0 \hspace{5mm}
         \forall \alpha.
         $$
         \end{description}
         \end{enumerate}
         The proof of Theorem 2 is given in Appendix B\@. Theorem 2 also
         shows the way of how algebras of perennials could be related
         to groups of symmetries: the generators of the group are
         locally Hamiltonian vector fields, and if they are
         globally Hamiltonian, they define perennials.

         The notion analogous to algebras of elementary perennials is
         that of {\em first-class canonical group}:

         {\bf Definition:} Let ${\cal G}$ be a group of
         transformations
         of $\Gamma$ satisfying the following requirements:
         \begin{enumerate}
         \item all elements of ${\cal G}$ are symmetries,
         \item all infinitesimal generators of ${\cal G}$ are
         globally Hamiltonian vector fields; let us denote by
         ${\cal PG}$ the algebra of the
         corresponding perennials; ${\cal PG}$ is a central extension
         (possibly trivial) of the Lie algebra ${\cal LG}$
         of the group $\cal G$; let ${\cal G}_c$ be the Lie group
         with the Lie algebra ${\cal PG}$; ${\cal G}_c$ is a central
         extension of $\cal G$,
          \item there is $\gamma
         \subset \tilde{\Gamma}$
         such that $\mbox{cl}({\cal G} \gamma) = \tilde{\Gamma}$
         (${\cal G}$
         acts
         almost transitively on $\tilde{\Gamma}/ \gamma$), where
         $\mbox{cl}
         ({\cal M})$ denotes the topological closure of the set ${\cal
         M}$,
         \item ${\cal N} = \{ \varphi \in {\cal G} \mid \varphi \gamma
         =
         \gamma, \hspace{2mm} \forall \gamma \}$ is a closed subgroup of
         ${\cal G}$; then it is easy to show that $\cal N$ is a normal
         subgroup of $\cal G$; moreover, ${\cal PN} = {\cal LN}$, as
         the perennial generators of $\cal PN$ can be chosen such that
         $o_{\mid \tilde{\Gamma}} = 0$.
         \end{enumerate}
         Then, the Lie group ${\tilde{\cal{G}}} = {\cal G}_c/{\cal N}$
         is called
         a first-class canonical group of the parametrized system.

         By the adjective ``first-class'', we distinguish this group
         from the canonical group of ref. [10]; the canonical group,
         whose elements need not be symmetries and which has to act
         transitively in $\Gamma$, can still be useful for constrained
         systems just for the methodical reasons.

         Such a first-class canonical group need not exist for an
         arbitrary parametrized system: because of condition 3,
         $\tilde{\Gamma}/\gamma$ must contain a dense subset which is
         a
         homogeneous space (of the form $\tilde{{\cal G}}/
         \tilde{{\cal G}}_0$,
         where
         $\tilde{{\cal G}}_0$ is a closed subgroup of $\tilde{{\cal G}}$).
         This problem must be further
         studied; here, we will suppose that a first-class canonical
         group exists for our system.

         Then, there will be many such groups, in general. The choice
         of the $\tilde{{\cal G}}$ is the freedom that we have in
         construction of the quantum theory. In particular, a chosen
         group $\tilde{{\cal G}}$ can have subgroups that satisfy all
         conditions for a first-class canonical group. We call such
         subgroups {\em first-class canonical subgroups}.

         The group quantization is finished by a choice of a unitary
         representation of the group $\tilde{{\cal G}}$ in a Hilbert
         space
         ${\cal H}$ satisfying the condition that all first-class
         canonical subgroups of $\tilde{{\cal G}}$ are represented
         irreducibly on ${\cal H}$. As a rule, there will be a number
         of such representations, and this gives the second freedom in
         the quantization.

         Let us close this section by a comparison of the group and
         algebraic quantizations as described above (see also [8]).
         Clearly, the Lie algebra of $\tilde{{\cal G}}$ defines a Lie
         algebra of classes of equivalent perennials because of
         condition 2. This Lie algebra satisfies property (i) of an
         algebra of elementary perennials of subsection 2.1,
         but not necessarily the orbit separation property (ii). Indeed,
         the separation property of the algebras and the transitivity
         (condition 3) of the groups do not have any direct relation
         in general. For example, if one quantizes systems with a
         discrete configuration space [27], then there are no algebras
         (canonical
         commutation rules), but one can quantize by the group method
         [28]; the group is not a Lie group, but a discrete one.
         Another example is given in ref. [29]: the canonical group
         $\tilde{{\cal G}}$ has to contain discrete elements similar
         to parity (these have to be represented by operators that
         are simultaneously unitary and self-adjoint, so their square
         is $id$) in order that it acts almost transitively. The Lie
         algebra of the component of identity of $\tilde{{\cal G}}$,
         however, defines a sufficient number of perennials so that
         the separation condition is satisfied.

         Another difference is that a given algebra of elementary
         perennials need not define a group (if the Hamiltonian
         vector fields are not complete in $\Gamma$) and, if it
         defines one, then the group can contain more information than
         the algebra. The former problem may be a spurious one: there
         are some suggestions that all ``quantizable'' variables
         should generate a group (see, e.g. [11] and the discussion
         about the Van Hove theorem in [26]). The latter difference is
         due to the fact that ``different'' Lie groups can have the
         ``same'' Lie algebras (example: $SO(2,1)$ and $Sl(2,\reell))$;
         then,
         not all representations of the Lie algebra are obtained from
         faithful representations of the group. This is, in fact,
         desirable, because it limits the second freedom in the
         quantization.

         The third difference is due to the ability of the algebras to
         incorporate relations like (9). These cannot be built into
         the group structure! Thus, the group can have representations
         which do not respect eq. (10). Then, algebraic quantization
         is more advantageous in such a case.

         It seems to follow that an obvious combination of both
         quantizations will be superior to any of them in general. In
         ref. [29], we show an example of such a quantization.

         \section{Transversal surface}

         The quantum mechanics constructed in the previous section
         have three remarkable features. 1) There is ``no Hamiltonian''
         (better, the Hamiltonian is equal zero), and all quantum
         observables are integrals of motion. We are faced by the
         problem to reconstruct changes and time dependence, which
         surely are observable features of our world, within these
         quantum mechanics. This is the problem of ``frozen dynamics''
         (see, e.g. [1], [2]). 2) For most systems, very few
         perennials are known (none for the general relativity),
         whereas our method needs even a complete system of perennials.
         3) There is  ``no gauge'' (better, no
         time function and gauge choice was necessary). This could
         surely be considered as ``an advantage'', but it is still of
         great interest to understand the relation of our quantum
         mechanics with those which are constructed via a choice of
         gauge. In fact, as a choice of gauge for a parametrized
         system has to do with time foliation ([1], [2]), and as
         perennials can be defined by their values at a given
         time, all three problems
         are related. In our theory, a gauge will be represented by a
         transversal surface.

         {\bf Definition}: Let $\Gamma_1$ be a $(2N-2\nu)$-dimensional
         surface
         in $\tilde{\Gamma} \subset \Gamma$ satisfying the following
         requirements.
         \begin{enumerate}
         \item Let $p \in \Gamma_1$ and $\gamma_p$ be the orbit
         through $p$; then
         $$
         T_p \Gamma_1 \oplus T_p \gamma_p = T_p \tilde{\Gamma}; \eqno(11)
         $$
         in particular, no non-zero vector is simultaneously tangential
         to both
         $\Gamma_1$ and $\gamma_p$.
         \item Let $\gamma$ be an arbitrary orbit, then $\gamma$
         intersects $\Gamma_1$ in at most one point.
         \end{enumerate}
         Then, $\Gamma_1$ is called {\em transversal surface}.

         Nice tool to work with $\Gamma_1$ is provided by two maps,
         $i_1$ and $\pi_1$, which are defined as follows. $i_1$ is the
         injection of $\Gamma_1$ into $\Gamma$; $i_1: \Gamma_1
         \rightarrow \Gamma$. $\pi_1$ is the projection from
         $\tilde{\Gamma}$ to $\Gamma_1$ given by
         $$
         \pi_1 p = \Gamma_1 \cap \gamma_p,
         $$
         where $\gamma_p$ is the orbit through $p\in \tilde{\Gamma}$.
         As $\Gamma_1 \cap \gamma_p$ may be empty, we have a
         nontrivial domain $\mbox{Dom} \pi_1 \subset \tilde{\Gamma}$.
         This subset of $\tilde{\Gamma}$ will play an important role,
         so we introduce a special name for it: {\em domain of $\Gamma_1$
         in} $\tilde{\Gamma}, \mbox{D}(\Gamma_1)$.

         We will call $\Gamma_1$ a {\em global transversal surface}, if
         $\mbox{D} (\Gamma_1) = \tilde{\Gamma}$, and {\em maximal
         transversal surface}, if there is no transversal surface
         $\Gamma_2$ such that $\mbox{D}(\Gamma_2)$ contains $\mbox{D}
         (\Gamma_1)$ as a proper subset.

         There are parametrized systems that do not admit any global
         transversal suface. This has to do with the so-called {\em
         global time problem} ([1], [2]), but is not identical to it.
         The global time problem has been studied for parametrized
         systems with a single constraint $C$ ([30], [25]); it arises,
         if there is no ``time function'' $T$. $T$ is a function on
         the phase space $\Gamma$ that grows along each orbit,
         $$
         \{ T, C\} >0.
         $$
         Hence, if there is such a time function, then $T =$ const.\
         will be a global transversal surface. However, the opposite
         is not true: there may be a global transversal surface, but
         no
         time function (if the orbits are closed). In the case of more
         constraints, some of them being pure gauges, the
         non-existence of global transversal surface can be due just
         to a Gribov problem for the gauge constraints, so that there
         is no problem with the time (parametrized SU(2) gauge
         theory). In the formalism which we are going to develop
         (following closely Dirac's ideas [5]), the global time and
         Gribov problem will turn up to be completely analogous and no
         advantage seems to result from treating them separately.

         The basic property of transversal surface is the following:

         {\bf Theorem 3}: Let $\Gamma_1$ be an arbitrary transversal
         surface, $i_1$ and $\pi_1$ the corresponding injection and
         projection, $\Omega_1 = i^\star_1 \Omega$ be the pull back of
         $\Omega$ to $\Gamma_1$ and $\tilde{\Omega}$ that of $\Omega$
         to $\tilde{\Gamma}$. Then,

         a) $(\Gamma_1, \Omega_1)$ is a symplectic manifold,

         b) $$\tilde{\Omega} = \pi^\star_1 \Omega_1
         \hspace{3mm} \mbox{on}
         \hspace{3mm} \mbox{D}(\Gamma_1). \eqno(12) $$

         The proof of Theorem 3 is given in Appendix B.

         The symplectic manifold $(\Gamma_1, \Omega_1)$ can be
         considered as a reduced phase space, if $\Gamma_1$ is a
         global transversal surface: the gauges are chosen and the
         constraints are solved. For any transversal surface
         $\Gamma_1$, the Poisson bracket on $\Gamma_1$ that is defined
         by $\Omega_1$ will be denoted by $\{\cdot, \cdot \}_1$.
         $\tilde{\Omega}$ is a pre-symplectic form on
         $\tilde{\Gamma}$; it is degenerate such that
         $$
         X_p = \{ X \in T_p \tilde{\Gamma} \mid \tilde{\Omega} (X,Y) =
         0 \hspace{3mm}, \forall \hspace{3mm} Y \in T_p \tilde{\Gamma} \}.
         $$
         Indeed, the distribution $X_p$ of vectors tangential to
         orbits is mapped to zero vector by $\pi_{1^\star}$.

         One would like, as next step, to quantize the system using a
         fixed transversal surface. This will be a quantization based
         on a ``choice of gauge''. Then, one will meet the so-called
         {\em multiple choice problem}: the quantum theories which
         will result from different choices $\Gamma_1$ and $\Gamma_2$ of
         transversal surface, will not be unitarily equivalent (see
         [1] and [2]). However, this inequivalence seems to be quite
         dependent from the method used. In this section, we describe
         a method which is ``gauge independent'' (solving, in a sense,
         the multiple choice problem).

         Let us consider first the algebraic method. For it we need a
         Lie algebra of elementary variables on $(\Gamma_1, \Omega_1)$.
         The crucial observation is that a given algebra of elementary
         perennials as defined in the previous section determines a
         unique Lie algebra of elementary variables with the same
         algebraical structure on $(\Gamma_1, \Omega_1)$. This can be
         seen as follows.

         First, we define a``projection'' of perennials to $(\Gamma_1,
         \Omega_1)$. Let $o$ be any perennial, and let $o_1$ be a
         function on $\Gamma_1$ given by
         $$
         o_1 = o\mid_{\Gamma_1} (= i^\star_1 o =  o \circ  i_1).
         $$
         In fact, $i^\star_1$ will map any function on $\Gamma$ to a
         function on $\Gamma_1$, but, for {\em perennials},
         $i^\star_1$ preserves {\em all} algebraic operations:

         {\bf Theorem 4}: Let ${\cal P}$ be the Poisson algebra of
         perennials on $\Gamma$, ${\cal J}_1$ its ideal of perennials
         vanishing at $\mbox{D}(\Gamma_1)$, and
         $$
         {\cal P}_1 = \{o_1 \in C^\infty (\Gamma_1) \mid
         o_1 = i^\star_1 o, o \in {\cal P} \}.
         $$
         Then,
         \begin{enumerate}
         \item ${\cal P}_1$ is a Poisson algebra on $(\Gamma_1,
         \Omega_1)$ (closed with respect to linear combination and
         product of functions and $\{ \cdot, \cdot \}_1$);
         \item $i^\star_1$ is a Poisson algebra homomorphism with
         kernel ${\cal J}_1$.
         \end{enumerate}
         The proof of Theorem 4 is given in Appendix B\@. In particular,
         if $\tilde{{\cal S}}$ is any algebra of elementary perennials,
         then $i^\star_1 \tilde{{\cal S}}$ is a well-defined Lie
         algebra of elementary variables on $(\Gamma_1, \Omega_1)$; the
         relations (like (9)) as well as the separation property will
         be preserved, only some new subalgebras of elementary
         variables might emerge if $\Gamma_1$ is not global.

         The algebraic structure itself is, however, not sensitive to
         the global properties of $\Gamma_1$: the full algebra of
         elementary perennials can be projected to an arbitrary small
         piece of transversal surface.

         The above projection procedure can be inverted.
         Suppose e.g. that $\Gamma_1$ is a global
         transversal surface and $o_1$ is an arbitrary
         $C^\infty$-function on $\Gamma_1$. Define $\tilde{o}:
         \mbox{D}(\Gamma_1) \rightarrow \komplex$ by
         $$
         \tilde{o} = o_1 \circ \pi_1 = \pi^\star_1 o.
         $$
         $\tilde{o}$ is a function on $\tilde{\Gamma}$ that is
         constant along orbits. Suppose that $\tilde{o}$ can be
         extended to a $C^\infty$-function on an open set that is
         dense in a neighbourhood of
         $\tilde{\Gamma}$ in $\Gamma$; if there is one such extension,
         $o$ say, then there will be many, and they will form an
         equivalence class of perennials. In general, there will be a
         complete set of functions on $\Gamma_1$ for which this
         construction
         can be performed. This procedure leads to a definition of
         perennials by their values at $\Gamma_1$. In most cases,
         one cannot calculate the perennials that are defined in
         this way as explicit functions on $\Gamma$. However, this
         is no problem: everything can be calculated from the
         pull-backs of the perennials to $\Gamma_1$ using the
         Theorem 4.

         Projection of symmetries is a more complicated business than
         that of perennials. Suppose $\varphi$ is a symmetry and
         $\Gamma_1$ a transversal surface. We define
         $$
         \varphi_1 = \pi_1 \circ \varphi \circ i_1.
         $$
         This is well-defined for any $p \in \Gamma_1$ only if
         $\varphi(p) \subset \mbox{D} (\Gamma_1)$, which in turn is
         equivalent
         to the condition that $\varphi$ preserves
         $\mbox{D}(\Gamma_1)$. Of course, any symmetry can be
         projected to a global transversal surface.

         Suppose that $\varphi$ and $\psi$ are two symmetries which
         preserve $\mbox{D}(\Gamma_1)$. Then, their composition $\psi
         \circ \varphi$ also does, and it is easy to prove that
         $$
         (\psi \circ \varphi)_1 = \psi_1 \circ \varphi_1.
         $$
         Indeed, if $\psi$ is a symmetry, then
         $$
         \pi_1 (\psi (i_1 (\pi_1 (p)))) = \pi_1 (\psi(p))
         $$
         as $i_1 (\pi_1(p))$ and $p$ lie at the same orbit
         and $\psi$ maps orbits onto orbits. Hence
         $$
         \psi_1 \circ \varphi_1 = (\pi_1 \circ \psi \circ i_1)\circ
         (\pi_1 \circ \varphi \circ i_1) =
         $$
         $$
         = (\pi_1 \circ \psi \circ i_1 \circ \pi_1) \circ \varphi
         \circ i_1 = \pi_1 \circ (\psi \circ \varphi) \circ i_1.
         $$
         Thus, the composition of maps is preserved by the projection
         only if the maps are symmetries.

         Suppose further that $\varphi$ is a symmetry preserving
         $\mbox{D}(\Gamma_1)$. Then, $\varphi_1$ is a symplectic map
         on $(\Gamma_1, \Omega_1)$. Indeed,
         $$
         \varphi^\star_1 \Omega_1 = i^\star_1
         (\varphi^\star (\pi^\star_1 \Omega_1)) = i^\star_1
         (\varphi^\star \tilde{\Omega})= i^\star_1 \tilde{\Omega}_1
         = \Omega_1;
         $$
         we have used Theorem 3b.

         Let ${\cal G}$ be a group of symmetries and ${\cal N}$ its
         normal subgroup that leaves orbits invariant. Let $\Gamma_1$
         be a transversal surface. Define
         $$
         {\cal G}_1 = \{ \varphi \in {\cal G} \mid \varphi \mbox{D}
         (\Gamma_1) \subset \mbox{D} (\Gamma_1) \},
         $$
         the subgroup of ${\cal G}$ leaving $\mbox{D}(\Gamma_1)$
         invariant. Clearly, ${\cal N}$ is a normal subgroup of ${\cal
         G}_1$. Denote by $a_1(\varphi)$ the projection of $\varphi
         \in {\cal G}$ to $\Gamma_1$. Then, the following theorem is
         an immediate consequence of the above considerations:

         {\bf Theorem 5}: ${\cal G}_1$ acts via $a_1$  as a group of
         symplectic diffeomorphisms on $(\Gamma_1, \Omega_1)$ and
         $$
         {\cal N} = \{ \varphi \in {\cal G}_1 \mid a_1 (\varphi) = id
         \}.
         $$

         We have seen that there are certain relations between group
         and algebraic quantization. This is based on a relation
         between infinitesimal generators of the group and perennials.
         This relation survives the projection to a transversal
         surface, as the following theorem shows.

         {\bf Theorem 6}: Let $\Gamma_1$ be a transversal surface and
         let $\varphi_t$ be a one-dimensional group of symmetries
         preserving $\mbox{D} (\Gamma_1)$ so that the projection
         $\varphi_{1t}$ of $\varphi_t$ to $\Gamma_1$ is well-defined.
         Let $X$ be the generator of $\varphi_t$ on $\Gamma$ and $X_1$
         that of $\varphi_{1t}$ on $\Gamma_1$. Let $p \in \Gamma_1, U$
         be a neighbourhood of $p$ in $\Gamma$ and $U \cap \Gamma_1 =
         U_1$ that of $p$ in $\Gamma_1$.  Let
         $f$ be a function in $U$ satisfying
         $$
         \langle \mbox{d}f,Y\rangle = \Omega(Y,X), \forall Y \in
         T_{p}\Gamma, \forall p \in U.
         $$
         Let, finally, $f_1$ be the projection of $f$ to $U_1$:
         $$
         f_1 = i^\star_1 f.
         $$
         Then,
         $$
         \langle \mbox{d}f_1,Y_1\rangle = \Omega_1(Y_1,X_1),
         \forall Y_1 \in T_{p}\Gamma_{1},\forall p \in U_1 .
         $$
         The proof of Theorem 6 is given in Appendix B. In particular,
         if $X$ is globally Hamiltonian, then $U= \Gamma$ and $f$ is a
         perennial whose projection $f_1$ to $\Gamma_1$ generates
         $\varphi_{1t}$.

         Suppose that $\tilde{{\cal G}}$ is a
         first-class canonical group of our system and $\Gamma_1$ is a
         global transversal surface. Then it is easy to show that
         $\tilde{\cal{G}}$ is also a canonical group of the system
         with the phase space $(\Gamma_1, \Omega_1)$
         (that is, canonical group in the sense of ref. [9], [10],
         [11]).

         In this way, any given group quantization of our parametrized
         system as defined in the previous section `induces'' a group
         quantization on the reduced phase space. The result of this
         quantization is independent of the choice of the transversal
         surface $\Gamma_1$ (it is a fixed representation of
         $\tilde{{\cal G}}$). What can be done in a case when there is
         no global transversal surface? A particular example is
         studied in ref. [29].

         The last theorem of this section concerns the relation
         between structures at different transversal surfaces.

         {\bf Theorem 7}: Let $\Gamma_1$ and $\Gamma_2$ be two
         transversal surfaces such that $\mbox{D} (\Gamma_1) \cap
         \mbox{D}(\Gamma_2) \neq \emptyset$. Let $\Omega_1$ and
         $\Omega_2$
         be the symplectic forms induced by $\Omega$ on $\Gamma_1$ and
         $\Gamma_2$, let $o_1$ and $o_2$ be the projections of a
         perennial $o$ to $\Gamma_1$ and $\Gamma_2$, and let
         $\varphi_1$ and $\varphi_2$ be those of a symmetry $\varphi$
         that satisfies $\varphi \mbox{D}(\Gamma_1) \subset
         \mbox{D}(\Gamma_1), \varphi \mbox{D}(\Gamma_2) \subset
         \mbox{D}(\Gamma_2)$. Let, finally, $\sigma: \Gamma_1
         \rightarrow \Gamma_2$ be defined by
         $\sigma(p) = \pi_2 (p) ,\forall p \in \Gamma_1$, and
         $\mbox{Dom}\sigma = \pi_1 (\mbox{D}(\Gamma_1) \cap \mbox{D}
         (\Gamma_2))$.
         Then:
         \begin{description}
         \item a) $\sigma^\star \Omega_2 = \Omega_1$,
         \item b) $\sigma^\star o_2 = o_1$,
         \item c) $\varphi_2 = \sigma \circ \varphi_1 \circ
         \sigma^{-1}$.
          \end{description}
          The proof of Theorem 7 is given in Appendix B.

         \section{Time evolutions and Hamiltonians}

         In ref. [5], the notion of a Hamiltonian of a system of
         interacting relativistic point particles is generalized:
         A Hamiltonian is any perennial that moves (via Poisson
         bracket) a given transversal surface around in the phase
         space. Accordingly, a) a system will have many Hamiltonians
         and b) each Hamiltonian is associated with a transversal
         surface (the same perennial can be a Hamiltonian for a
         transversal surface $\Gamma_1$, but no Hamiltonian for
         another surface $\Gamma_2$). In this way, conditions on
         Hamiltonians are weakened on one hand; although this
         weakening
         was developed by Dirac for other purposes, it seems to be
         very useful for the theory of general parametrized systems
         with their problem of time. On the other hand, the conditions
         are stronger in the sense, that such a Hamiltonian always
         generates a {\em symmetry} group of the system.

         One of the building blocks used by Dirac to construct a time
         evolution is a particular transversal surface. (His ``three
         forms of relativistic dynamics'' are associated with the
         three most symmetrical ``transversal'' surfaces in Minkowski
         spacetime: the spacelike plane and hyperboloid and the null
         plane.) It seems that a time evolution of an isolated
         parametrized system cannot be constructed completely just by
         means of the system's own perennials. This can quite
         convincingly be demonstrated by studying the dynamics of a
         particle in Minkowski spacetime (see, e.g. ref. [13]). That
         is one of the reasons for distinguishing between
         ``perennials'' and ``observables''. However, it is hoped that
         the ``non-perennial'' observable aspects can be obtained by
         means of perennials of some suitably extended system; for
         example, one could try to couple gravity to some matter
         clock, etc., see, e.g. ref. [31].

         In this section we generalize and further develop Dirac's
         ideas on Hamiltonians and time evolution. The crucial notions
         are those of ``time level'' and of ``the same measurement at
         different times''. Indeed, to observe a ``change'' in a
         system between the ``time levels'' $t_1$ and $t_2$, one has
         to perform two ``equal measurements'', one at $t_1$ and one
         at $t_2$, and compare their results. This idea can be made
         precise as follows.

         Choose a transversal surface $\Gamma_0$ with domain
         $\mbox{D}(\Gamma_0)$ in $\tilde{\Gamma}$ and a
         one-dimensional group of symmetries $h(t)$ preserving
         $\mbox{D}(\Gamma_0)$. Let the generator of $h(t)$ be the
         perennial $h$. Define
         $$
         \Gamma_t = h(t) \Gamma_0.
         $$
         $\Gamma_t$ is a transversal surface with the same domain as
         $\Gamma_0$, because $h(t)$ is a symmetry presenting $\mbox{D}
         (\Gamma_0)$. The one-dimensional family $\{\Gamma_t\}$ sweeps
         a $(2N-2\nu+1)$-dimensional surface $\tilde{\Gamma_0}$ within
         the $(2N-\nu)$-dimensional constraint surface
         $\tilde{\Gamma}$.
         We will call the surfaces $\Gamma_t$ in $\tilde{\Gamma}$ {\em
         time levels}. Observe that the time levels cover only a small
         portion of the constraint hypersurface $\tilde{\Gamma}$ if
         there is more than one constraint $(\nu >1)$.

         Suppose next that we perform a measurement at the time level
         $\Gamma_0$. Let this be a measurement of the value $o(p)$ of
         a perennial $o$ at some state $p \in \Gamma_0$ of the system
         at $t = 0$. What is the corresponding, or ``the same'',
         measurement at the time level $\Gamma_t$? We define: it is
         the measurement of the value of the {\em time-shifted}
         perennial $o_t$, given by
         $$
         o_t = h^\star (-t) o = o \circ h(-t),
         $$
         at $\Gamma_t$. At least formally, this definition is
         reasonable because of the following theorem.

         {\bf Theorem 8}: Let $o$ and $o'$ be two perennials and
         $\varphi$ a symmetry. Then, $\varphi^\star o$ and
         $\varphi^{\star} o'$ are again perennials, and the following
         relations hold
         $$
         \alpha \varphi^\star o + \beta \varphi^\star o' =
         \varphi^\star (\alpha o + \beta o'),
         $$
         $$
         \varphi^\star o \cdot \varphi^\star o' = \varphi^\star (o
         \cdot o'),
         $$
         $$
         \{ \varphi^\star o, \varphi^\star o'\} = \varphi^\star \{
         o,o'\}.
         $$
         The proof is very simple and we leave it to the reader. Thus,
         a symmetry preserves the perennial property and the Poisson
         algebra of perennials. Theorem 8 enables us to define a
         complete system of related (``the same'') measurements at any
         two different time levels. This provides a formal reason why
         the time evolution should be defined by a symmetry group. The
         one-dimensional family $\{ o_t\}$ of time-shifted perennials
         can be also considered as a kind of ``evolving constant of
         motion'' (cf. [18]). On one hand, however, these evolving
         constants need additional perennials to be well-defined,
         whereas Rovelli's supply additional perennials, on the other,
         they are well-defined for any dimension, whereas
         Rovelli's definition seems to work only for one-dimensional
         orbits.

         The notion of time-shifted perennial enables us to study the
         motion of the system is a quantitative way. First, we
         introduce an {\em auxiliary rest frame}. This frame is formed by
         the orbits of the group $h(t)$ in $\tilde{\Gamma_0}$. That
         is, the points $p \in \Gamma_0$ and $h(t) p \in \Gamma_t$ are
         considered as ``the same''. Indeed, any two ``equal''
         measurements, of $o$ and of $h^\star(-t)o$, will give the
         same results at $p$ and $h(t)p$:
         $$
         o(p) = o \circ h(t) (h(-t) p). \eqno(13)
         $$

         To study the classical solutions -- the possible motions of
         the system -- relatively to the auxiliary rest frame, we have
         to represent the constraint orbits by curves in
         $\tilde{\Gamma_0}$. This is straightforward: Let $\gamma$ be
         an orbit in $\mbox{D}(\Gamma_0)$ and set $\eta_\gamma(t)$ be
         the intersection point of $\Gamma_t$ and $\gamma$:
         $$
         \eta_\gamma (t) = \Gamma_t \cap \gamma;
         $$
         $\eta_\gamma(t), t \in \reell$ is desired representation as
         the point $\eta_\gamma(t) \in \gamma$ is well-defined for
         each $t \in \reell$. Indeed, $\Gamma_t$ is transversal, so
         $\Gamma_t$ intersects $\gamma$ at one point at most. $\gamma$
         intersects $\Gamma_0$ as $\gamma \in \mbox{D}(\Gamma_0)$.
         But $h(t)$ is a symmetry preserving $\mbox{D}(\Gamma_0)$, so
         $\mbox{D}(h(t) \Gamma_0) = \mbox{D}(\Gamma_0)$, and $\gamma$
         must also intersect $\Gamma_t$. From the definition of
         $\eta_\gamma(t)$, it follows that
         $$
         \eta_\gamma (\reell) = \tilde{\Gamma}_0 \cap \gamma.
         $$

         Now, we can define: the time evolution is the
         change in the results of one and the same measurement done
         at different times along the dynamical trajectory of the
         system. These results are given by the numbers
         $o_t(\eta_\gamma(t))$. The number
         $o_t(\eta_\gamma(t))$ can be expressed in two different
         ways. First, from the definition of $o_t$, it follows that
         $$
         o_t(\eta_\gamma(t)) = o(\xi_\gamma(t)) = \tilde{o}
         (\xi_\gamma(t)),
         $$
         where $\tilde{o}$ is the projection of the perennial $o$ to
         $\Gamma_0$ and
         $$
         \xi_\gamma(t) = h(-t) \eta_\gamma(t). \eqno(14)
         $$
         $\xi_\gamma(t)$ is the projection of the curve
         $\eta_\gamma(t)$ down to $\Gamma_0$ by means of the auxiliary
         rest frame. This means that $\xi_\gamma(t)$ describes the
         motion $\eta_\gamma(t)$ relatively to the rest frame. Second,
         it follows from the constancy of $o_t$ along orbits that
         $$
         o_t (\eta_\gamma(t)) = o_t (\eta_\gamma (0))
         $$
         as both points $\eta_\gamma(t)$ and $\eta_\gamma(0)$ lie at
         the same orbit $\gamma$. In this way, we have obtained the
         ``(classical) Schr\"{o}dinger'' and the ``(classical) Heisenberg
         picture of the dynamics''. Within
         the former, there is a time-independent observable $\tilde{o}
         = o\mid_{\Gamma_0}$ of the reduced system, and a
         time-dependent state $\xi_\gamma(t) \in \Gamma_0$ of the
         system. Within the latter, there is a time-dependent
         observable $\tilde{o}_t = o_t \mid_{\Gamma_0}$ of the reduced
         system, and a time-independent state $\xi_\gamma(0) \in
         \Gamma_0$ of the system. The result of the measurement at
         time $t$ is given by
         $$
         \tilde{o}_0 (\xi_\gamma(t)) = \tilde{o}_t (\xi_\gamma
         (0)).
         $$
         The time-dependence of the measurement results can be
         calculated using the following theorem.

        {\bf Theorem 9}: Let $(\Gamma_0, \Omega_0)$ be the
         symplectic space with $\Omega_0 = i_0^\star \Omega$, and let
         $H:\Gamma_0 \rightarrow \reell$ be defined by
         $$
         H = - h\mid_{\Gamma_0}.
         $$
         Then:

         A) The curve $\xi_\gamma(t)$ defined by (14) for each
         $\gamma \subset \mbox{D}(\Gamma_0)$ is an integral curve of
         the Hamiltonian vector field of the function $H$ in
         $(\Gamma_0, \Omega_0)$.

         B) The one-dimensional family $\{\tilde{o}_t \}$ of
         observables satisfies the equation
         $$
         \frac{d \tilde{o}_t}{dt} = \{\tilde{o}_t, H \}_0. \eqno(15)
         $$
         The proof of the Theorem 9 is given in the Appendix B.

         Theorem 9 shows that everything about time evolution can be
         calculated from the pull-backs of the relevant perennials
         to a transversal surface.

         There are clearly some reasons to call the function $H$ in
         Theorem 9 a
         {\em Hamiltonian}. However, there is still a problem: the
         curve $\eta_\gamma(t)$ at $\gamma$ need not represent any
         non-trivial motion. Indeed, $\eta_\gamma(t)$ represents some
         part of the maximal classical solution $({\cal M}_\gamma,
         Q_\gamma)$ associated with $\gamma$. Let us denote this part
         by $({\cal M}(\Gamma_0, h, \gamma), Q (\Gamma_0, h
         ,\gamma))$.
         If ${\cal M} (\Gamma_0, h, \gamma)$ does not contain any open
         subset of ${\cal M}_\gamma$, then $\eta_\gamma(t)$ represents
         just a one-dimensional set of gauge-equivalent initial data.
         Then, there is no reason to call the corresponding function
         $H$ a Hamiltonian. If $\eta_\gamma (t)$ {\em is} a
         non-trivial part of a classical solution, then we can
         distinguish the following two cases:

         1. Let $\Gamma_0$ be a global transversal surface and let for
         each orbit $\gamma \in \tilde{\Gamma}, {\cal M} (\Gamma_0, h,
         \gamma) = {\cal M}_\gamma$. Then, we call $H$ a {\em
         Hamiltonian associated with $\Gamma_0$}.

         2. There is an orbit $\gamma \in \mbox{D}(\Gamma_0)$ such
         that ${\cal M}(\Gamma_0, h, \gamma)$ contains an open proper
         subset ${\cal M}_\gamma$. Then, $H$ is called a {\em partial
         Hamiltonian associated with $\Gamma_0$}.

         In general, there will be only partial Hamiltonians. In each
         of the two cases, however, $\eta_\gamma(t)$ represents a
         non-trivial piece of a maximal classical solution at least
         within an open subset of orbits in $\tilde{\Gamma}/\gamma$.
         That is, our construction determines a foliation and a gauge
         choice in some open subset of maximal classical solutions;
         these foliations will, however, cover only a part of each
         solution in general. One can hope to improve this
         unsatisfactory situation by a choice of a whole family of
         partial Hamiltonians such that, starting from one transversal
         surface $\Gamma_0$, all points of all maximal classical
         solutions of $\mbox{D} (\Gamma_0)$ will be covered. Such a
         family of partial Hamiltonians associated with $\Gamma_0$
         will be called {\em complete}. A complete family of partial
         Hamiltonians will exist if the group $\cal G$ of symmetries
         acts transitively (or almost transitively) on the constraint
         manifold of the system.

         Consider $\Gamma_0$ which is a global transversal surface and
         complete family of partial Hamiltonians associated with
         $\Gamma_0$. Then, the family of the corresponding time levels
         can be parametrized by $(t_1, \ldots, t_k) \in \reell^k$,
         where $k$ is the number of partial Hamiltonians in the family.
         This corresponds to a $k$-parameter family of time slices
         inside of each solution. Thus, if we apply these ideas to
         general relativity, we obtain a $k$-dimensional subset of the
         full system of ``many-finger time'' slices in solutions. In
         this way, our method leads to a sort of compromise between a
         fixed choice of time foliation for each spacetime (i.e., the
         constant mean external curvature hypersurfaces, cf. ref.
         [32]) and a complete functional time formalism that
         would enable one to calculate the evolution between {\em any}
         two spacelike hypersurfaces in any spacetime (cf. ref. [1]).

         As for the quantum mechanics, the construction of the
         corresponding Schr\"{o}dinger and Heisenberg pictures of the
         dynamics is straightforward, because all perennials and
         symmetry groups needed for that aim possess already their
         operator representations.

         Indeed, let $\tilde{\cal{G}}$ be a first-class canonical
         group of our system and $\cal H$ a representation Hilbert
         space with the scalar product $(\cdot ,\cdot )$. Let the
         representative of $g \in \tilde{\cal{G}}$ be the unitary
         operator $\hat{U}_g$  and the
         representative of a perennial $o \in \cal{L}\tilde{\cal G}$
         be $\hat{o}$. We interpret the elements of $\cal{H}$ as
         Heisenberg states and the operators like $\hat{o}$ as
         Schr\"{o}dinger observables of the system.

         Suppose that $h(t)$ is a one-dimensional subgroup of
         $\tilde{\cal{G}}$ defining, together with some transversal
         surface $\Gamma_0$, a time evolution of our system. Let
         $\hat{U} (t)$ be the representatives of $h(t)$. Then,
         $\hat{U} (t)$ is interpreted as the unitary time evolution
         operator. We define a Schr\"{o}dinger state $\psi_t$ of
         the system by
         \[
           \psi_t = \hat{U} (t), \forall \psi \in \cal{H},
         \]
         and a Heisenberg observable $\hat{o}_t$ of the system by
         \[
           \hat{o}_t = \hat{\cal{U}}^{-1}(t) \hat{o} \hspace{1mm}
           \hat{\cal{U}}(t).
         \]

         Using these definitions together with the Theorems 4 and 9
         enables us to show that the ``perennial
         formalism'' method of quantization gives the same results
         as other methods for systems whose quantum mechanics is
         well-known. Let us briefly show an example. Let $q^{1},
         \cdots ,q^{n}, p_{1},\cdots ,p_{n}$ be the canonical
         coordinates of a (non-constrained) system with the phase
         space $\Gamma
         \cong \reell^{2n}$ and let $H = H(q,p)$ be the Hamiltonian
         of the
         system. Let us parametrize the system by extending $\Gamma$
         by two dimensions, adding the coordinates $q^0$ (originally
         the time) and $p_0$ (the conjugate momentum to $q^0$). Then
         the parametrized system with the constraint
         \[
         C \equiv p_0 + H(q^{1},\cdots ,q^{n},p_{1},\cdots ,p_{n})
         \]
         will be equivalent to the original system. Let us quantize
         this parametrized system by the ``perennial formalism''
         method.

         Clearly, $\Gamma_0$ defined by $q^0 = 0,C = 0$ is a global
         transversal surface. Let us define the perennials $Q^{1},
         \cdots ,Q^{n}$ and $P_{1},\cdots ,P_{n}$ by their
         pull-backs to $\Gamma_0$ as follows
         \[
         Q^i|_{\Gamma_0} = q^i ,P_i|_{\Gamma_0} = p_i ,i = 1,\cdots ,n.
         \]
         There is another useful perennial: it is $p_0$ (the
         Hamiltonian is independent of $q^0$). $p_0$ satisfies all
         our conditions on Hamiltonian. Further, we obtain that
         \[
         -p_0|_{\Gamma_0} = H(q^{1},\cdots ,q^{n},p_{1},\cdots ,p_{n}).
         \]
         Thus, the pull-backs of the perennials $Q^{1},\cdots ,Q^{n},
         P_{1},\cdots ,P_{n}$ and $p_0$
         have the same Lie algebra as the original variables $q^{1},
         \cdots ,q^{n},p_{1},\cdots , p_{n}$ and $H$. Then, one of
         the possible representations of this algebra of elementary
         perennials coincides with the original quantum mechanics.
         Observe that (with the exception of $p_0$) none of the
         perennials is known explicitly.

         {\bf Acknowledgement} Useful discussions with R. Beig, B.
         Br\"{u}gmann, D. Giulini, A. Higuchi, C.J. Isham, B.S. Kay,
         J. Tolar and L. Ziewer are thankfully acknowledged. The
         author thanks for hospitality by the Workshop on
         Mathematical Relativity at the Erwin Schr\"{o}dinger
         International Institute for Mathematical Physics in Vienna
         and by the Theoretical Physics Group at the Blackett
         Laboratory of the Imperial College in London, where parts
         of the paper have been written. The work was supported in
         part by Schweizerischer Nationalfonds, The Workshop on
         Mathematical Relativity in Vienna and by the Theoretical
         Physics Group at the Imperial College, London.

         \section*{Appendix A: System of $\nu$ free relativistic
         particles}

         The system of $\nu$ free relativistic massive particles is
         the simplest example of a parametrized system with several
         constraints. The configuration space $C$ is $ \reell^N, N =
         4\nu$ with coordinates $x^a_\alpha, \alpha = 1, \cdots,\nu$,
         $a = 0,1,2,3$, which are defined by an inertial system
         in Minkowski spacetime. The phase space is $\Gamma
         \cong \reell^{2N}$, and the canonical
         coordinates can be chosen as $x^a_\alpha$ and $p_{\alpha a}$.
         The action in the form (1) can be written as
         $$
         S = \int d\tau \sum_{\alpha=1}^{\nu}(p_{\alpha a}
         \dot{x}^a_\alpha - N_\alpha C_\alpha),
         $$
         where the constraints $C_\alpha$ are defined by
         $$
         C_\alpha = \frac{1}{2} (\eta^{ab} p_{\alpha a} p_{\alpha
         b} + m^2_\alpha),
         $$
         where $\eta^{ab}$ is the Minkowski metric and $m_\alpha$
         is the mass of $\alpha$-th particle. The constraint surface
         $\tilde{\Gamma}$ consists of $2^\nu$ components,
         $\tilde{\Gamma}_+$ is that one at which all $p_{\alpha0}$ are
         negative.

         The orbits are maximal integral manifolds of the vector fields
         $$
         \dot{x}^a_\alpha = \eta^{ab} p_{\alpha b}, \hspace{5mm}
         \dot{p}^a_\alpha = 0.
         $$
         Thus the vector fields are constant along each orbit and so
         their integral can be written immediately
         $$
         x^a_\alpha = X^a_\alpha - \zeta_\alpha \eta^{ab} P_{\alpha b},
         \eqno(A1a)
         $$
         $$
         p_{\alpha b} = P_{\alpha b}, \eqno(A1b)
         $$
         where $\zeta_\alpha$ are arbitrary real parameters and
         $X^a_\alpha$, $P_{\alpha a}$ is a set of $8\nu$ constants
         satisfying the constraints
         $$
         \eta^{ab} P_{\alpha a} P_{\alpha b} + m^2_\alpha = 0,
         \hspace{3mm} \alpha = 1, \cdots, \nu.
         $$
         Thus, they are $\nu$-dimensional planes, and $\zeta_\alpha$
         can be considered as coordinates on them.

         A solution curve is determined by eqs. (8)
         $$
         \dot{x}^a_\alpha = N_\alpha \eta^{ab} p_{\alpha b},
         $$
         $$
         \dot{p}_{\alpha a} = 0.
         $$
         If the Lagrange multipliers are given functions of the
         parameter $\tau$, then we have
         $$
         \zeta_\alpha(\tau) = - \int^{\tau}_{0} d\tau' N_\alpha(\tau').
         $$

         Classical solutions can be considered as maps $\sigma$ of $\nu$
         real lines $\reell$  into Minkowski spacetime. Thus, the bare
         manifold ${\cal M}$ is $\stackrel{\nu}{\cup}_{\alpha=1}
         \reell_\alpha$, where $\reell_\alpha = \reell, \hspace{2mm}
         \forall \hspace{2mm}
         \alpha$, and the fields are given by points of the Minkowski
         space $V$, as described by coordinates $x^a$. Hence, $\sigma$
         can be described by $4 \nu$ functions $\sigma^a_\alpha$ as
         follows
         $$
         x^a (\tau_\alpha) = \sigma^a_\alpha (\tau_\alpha),
         \tau_\alpha \in \reell_\alpha; \eqno(A2)
         $$
         $x^\alpha$ are scalar fields on ${\cal M}$.

         Let $\sigma: {\cal M} \rightarrow V$ and $\sigma' : {\cal M}
         \rightarrow V$ be two solutions. We say that $\sigma'$ is
         part of $\sigma$, if there is a map
         $$
         \varphi: {\cal M} \rightarrow {\cal M}
         $$
         such that $\varphi (\mbox{Dom} \sigma') \subset \mbox{Dom}
         \sigma$ and
         $$
         \sigma' = \sigma \circ \varphi.
         $$
         $\varphi$ is called a {\em reparametrization}.

         Any curve at the orbit (A1) can be given by
         $$
         \zeta_\alpha = f_ \alpha (\tau), \eqno(A3)
         $$
         where $f_\alpha (\tau)$ are piecewise smooth real functions.
         Then, the corresponding solution $\sigma$ is given by
         $$
         \sigma^a_\alpha(\tau_\alpha) = X^a_\alpha - f_\alpha
         (\tau_\alpha) \eta^{ab} P_{\alpha b}. \eqno(A4)
         $$
         This is the relation between a curve at an orbit and the
         corresponding classical solutions.

         The maximal classical solution $({\cal M}_\gamma, Q_\gamma)$ for the
         orbit $\gamma$ given by eq. (A1) can be represented by the
         curve with
         $$
         f_\alpha (\tau) = \tau,
         $$
         so that the corresponding map $\sigma_\gamma$ is
         $$
         \sigma^a_{\gamma\alpha} (\tau_\alpha) = X^a_\alpha -
         \tau_\alpha \eta^{ab} P_{\alpha b}.
         $$
         Indeed, consider any other curve at $\gamma$; it is
         represented by eq. (A3) and the corresponding solution is
         given by (A4). Consider the map $\varphi\;:\; {\cal M}
         \rightarrow {\cal M}$ defined by
         $$
         \varphi (\tau_\alpha) = f_\alpha (\tau_\alpha).
         $$
         As $f_\alpha$ are real functions, we have $f_\alpha
         (\tau_\alpha) \in \reell_\alpha \hspace{3mm}\forall\hspace{3mm}
         \alpha, \tau_\alpha \in \reell_\alpha$,
         and
         $$
         \sigma = \sigma_\gamma \circ \varphi.
         $$
         Thus, $\sigma$ is a part of $\sigma_\gamma$.

         An example of transversal surface $\Gamma_1$ is given by
         $$
         x^0_\alpha = 0, \eqno(A5a)
         $$
         $$
         C_\alpha = 0 \hspace{3mm} \forall \alpha. \eqno(A5b)
         $$
         $\Gamma_1$ is a $6\nu$-dimensional surface in $\tilde{\Gamma}$.
         We must show that the two conditions on transversal surface
         are fulfilled.
         \begin{enumerate}
         \item Any tangential vector to the orbit $\gamma$ given by
         eqs. (A1) has the form
         $$
         \dot{x}^a_\alpha = N_\alpha \eta^{ab} P_{\alpha b},
         $$
         $$
         \dot{p}_{\alpha a} = 0,
         $$
         where
         $$
         P_{\alpha 0} = \pm \sqrt{P^2_{\alpha 1} + P^2_{\alpha 2} +
         P^2_{\alpha 3} + m^2_\alpha} \neq 0.
         $$
         It will be tangential to $\Gamma_1$, if $\dot{x}^0_\alpha =
         0$ for all $\alpha$, but this implies $N_\alpha = 0,
         \hspace{2mm} \forall \alpha$, or $\dot{x}^a_\alpha = 0,
         \dot{p}_{2 a} = 0$.
         \item The point of intersection of the orbit (A1) with the
         surface (A5) is given by the solution of the quations
         $$
         X^0_\alpha - \zeta_\alpha \eta^{0b} P_{0 b} = 0.
         $$
         This equation has always a unique solution
         $$
         \zeta_\alpha = - \frac{X^0_\alpha}{P_{\alpha0}}. \eqno(A6)
         $$
         \end{enumerate}
         Thus, $\Gamma_1$ is even globally transversal.

         Let the coordinates on $\Gamma_1$ be $(y^k_\alpha,
         q_{\alpha k})$, $\alpha = 1, \cdots, \nu , k = 1,2,3$, such
         that the imbedding map $i_1$ is given by the equations
         $$
         \begin{array}{lll}
         x^0_{\alpha} & = 0, & x^k_\alpha = y^k_\alpha,  \nonumber \\
         p_{\alpha0} &= \pm \sqrt{q^2_{\alpha1} + q^2_{\alpha2} +
         q^2_{\alpha3} + m^2_\alpha}, & p_{\alpha k} =
         q_{\alpha k}.\end{array}
         $$
         The pull back $\Omega_1$ of $\Omega$ is given in these
         coordinates by
         $$
         \Omega_1 = \sum_{\alpha}dq_{\alpha k}  \wedge dy^k_\alpha.
         $$
         The map $\pi_1$ is defined by eq. (A6): all points of the
         orbit (A1) are mapped by $\pi_1$ to the point
         $$
         x^0_\alpha = 0, \hspace{3mm} x^k_\alpha = X^k_\alpha +
         X^0_\alpha \frac{P_{\alpha k}}{P_{\alpha0}},
         $$
         $$
         p_{\alpha a} = P_{\alpha a}.
         $$

         Finally, we give several examples of Hamiltonians and
         partial Hamiltonians.

         \noindent {\bf Example 1}:
         $$
         h = \sum_{\alpha} p_{\alpha 0}.
         $$
         $h$ is clearly a perennial. The group $h(t)$ generated by $h$
         contains the transformations:
         $$
         x^0_\alpha \rightarrow x^0_\alpha + t,
         $$
         $$
         x_\alpha^k \rightarrow x^k_\alpha \hspace{2mm}, \hspace{2mm}
         p_{\alpha a} \rightarrow p_{\alpha a}.
         $$
         Thus, $\Gamma_t$ is given by
         $$
         x^0_\alpha = t \hspace{2mm}, \hspace{2mm} C_\alpha = 0,
         \hspace{3mm} \forall \alpha
         $$
         (we rename $\Gamma_1$ to $\Gamma_0$). The point
         $\eta_\gamma(t)$ of intersection of $\Gamma_t$ with the orbit
         $\gamma$ given by eq. (A1) is determined by the equation
         $$
         X^0_\alpha + \zeta_\alpha P_{\alpha 0} = t.
         $$
         Hence, the curve $\eta_\gamma(t)$ is defined by
         $$
         \zeta_\alpha (t) = \frac{t-X^0_\alpha}{P_{\alpha0}}.
         $$
         This curve represents the maximal solution $\sigma_\gamma$,
         as the reparametrization
         $$
         t'_\alpha = \frac{t_\alpha - X^0_\alpha}{P_{\alpha 0}}
         $$
         is invertible.

         Hence, $ H = - h \mid_{\Gamma_o} = \pm \sum_{\alpha}
         \sqrt{q^2_{\alpha 1}+ q^2_{\alpha 2} + q^2_{\alpha 3}+
         m^2_\alpha}$ is a Hamiltonian. The projection
         $ \xi_\gamma(t)$ of $\eta_\gamma(t)$
         to $\Gamma_0$ is
         $$
         y^k_\alpha = X^k_\alpha - \frac{P_{\alpha k}}{P_{\alpha 0}} (t
         - X^0_\alpha),
         $$
         $$
         q_{\alpha k} = P_{\alpha k};
         $$
         this is just the motion of $\nu$ particles in the reduced
         phase space.

         \noindent {\bf Example 2}:
         $$
         h = \sum_{\alpha} p_{\alpha3}.
         $$
         $h(t)$ is the translation in 3rd direction:
         $$
         x^a_\alpha \rightarrow x^a_\alpha + \delta^a_3 t
         \hspace{2mm}, \hspace{2mm} p_{\alpha a} \rightarrow p_{\alpha
         a}.
         $$
         The time surfaces $\Gamma_t$:
         $$
         \Gamma_t = h(t) \Gamma_0 = \Gamma_0.
         $$
         The point $\eta_\gamma(t)$ of intersection between the orbit
         $\gamma$ given by eq. (A1) and $\Gamma_t$ has the coordinates
         $$
         \zeta_\alpha (t) = - \frac{X^0_\alpha}{P_{\alpha0}}.
         $$
         Thus, the curve $\eta_\gamma(t)$ consists of just one point.
         The coordinates of this point in $\Gamma$ are
         $$
         x^a_\alpha =  X^a_\alpha + X^0_\alpha \eta^{ab}
         \frac{P_{\alpha b}}{P_{\alpha 0}},
         $$
         $$
         p_{\alpha a} = P_{\alpha a}.
         $$
         This is just one initial data for the maximal solution
         $\sigma_\gamma$ at the point
         $$
         t_\alpha = - \frac{X^0_\alpha}{P_{\alpha 0}}
         $$
         of ${\cal M}_\gamma$. Thus, ${\cal M}(\Gamma_0, h, \gamma)$
         is one point in ${\cal M}_\gamma$ and does not contain
         an open subset of ${\cal M}_\gamma : h$ is not suitable to
         define a Hamiltonian for $\Gamma_0$.

         \noindent {\bf Example 3}:
         $$
         h = p_{10}.
         $$
         $\Gamma_t$ is given by the equations
         $$
         x^a_\alpha = \delta^1_\alpha \delta^a_0 t \hspace{2mm},
         \hspace{2mm} C_\alpha = 0, \, \, \forall\, \,\alpha.
         $$
         The intersection point $\eta_\gamma(t)$ has the coordinates
         $$
         \zeta_1(t) = \frac{t-X^{01}}{P_{10}},
         $$
         $$
         \zeta_\alpha(t) = - \frac{X^0_\alpha}{P_{\alpha0}}, \hspace{3mm}
         \alpha \neq 1.
         $$
         The corresponding solution can be mapped in the maximal
         solution by the reparametrization $\varphi$ that is given by
         $$
         t'_1 = \frac{t-X^0_1}{P_{10}} \hspace{2mm} , t'_\alpha =
         \frac{X_\alpha}{P_{\alpha 0}} \hspace{2mm} ,
         \hspace{2mm}\alpha \neq 1.
         $$
         The range of $\varphi$ is an open proper subset of ${\cal
         M}_\gamma$. It follows that
         $$
         H_1 = - h \mid_{\Gamma_0} = \pm
         \sqrt{q^2_{11}+q^2_{12}+q^2_{13} + m^2_1}
         $$
         is a partial Hamiltonian.

         Clearly, the $\nu$ partial Hamiltonians
         $$
         H_\alpha = \pm \sqrt{q^2_{\alpha 1} +q^2_{\alpha 2}
         q^2_{\alpha 3}+ m^2_\alpha},\alpha = 1,\cdots ,\nu ,
         $$
         are sufficient to move from any initial data to any other
         initial data of any maximal solution.

         \section*{Appendix B: Proofs}

         \subsection*{B1. Proof of the Theorem 2}

         1) Let $p \in \tilde{\Gamma} \cap \mbox{Dom} \varphi$ and $q
         = \varphi p$. Then there is a neighbourhood $U$ of $p$ in
         $\tilde{\Gamma}$ such that $\varphi U $ is a neighbourhood of
         $q$ in $\tilde{\Gamma}$, because $\varphi \tilde{\Gamma}
         \subset \tilde{\Gamma}$. It follow that
         $$
         \varphi_\star T_p \tilde{\Gamma} = T_q \tilde{\Gamma}
         $$
         and
         $$
         (\varphi^{-1})^\star N_p = N_q.
         $$
         As $X_p = \Omega^{-1} N_p$, $X_q = \Omega^{-1} N_q$ and
         $\varphi$ preserves $\Omega$, we obtain also that
         $$
         \varphi_\star X_p = X_q.
         $$
         Thus, the distribution $X_p$ is preserved by $\varphi$. This
         implies that the orbit through $p$ is mapped by $p$ into the
         orbit through $q$. However, $\varphi$ has an inverse; for an
         analogous reason, this inverse must map the orbit through $q$
         into the orbit through $p$. Hence, the map is onto.

         2) The claim a) follows directly from the definition
         ($\varphi$ preserves $\tilde{\Gamma}$).

         b) $\xi^A$ is an infinitesimal symplectic map,
         $$
         {\cal L}_\xi \Omega = 0,
         $$
         and the existence of $U$ and $f$ is a well-known property of
         such maps (see, e.g. [2]). Finally,
         $$
         \{ f, C_\alpha\}_{\tilde{\Gamma}}  = (\Omega^{AB} \partial_A
         f
         \partial_B C_\alpha)_{\tilde{\Gamma}} = - (\xi^B\partial_B
         C_\alpha)_{\tilde{\Gamma}} = 0
         $$
         because of 2a), QED.

         \subsection*{B2. Proof of Theorem 3}

         a) The form $\Omega$ is closed, so $i^\star_1 \Omega$ is also
         closed; we have, therefore, to show that $i^\star_1 \Omega$
         is non-degenerate. The pull-back $\tilde{\Omega}$ of
         $\Omega$ to ${\tilde{\Gamma}}$ {\em is} degenerate, it is
         well-known that
         $$
         X_p = \{X \in T_p \tilde{\Gamma} \mid \Omega(X,Y) = 0,
         \hspace{2mm} \forall \hspace{2mm} Y \in T_p\tilde{\Gamma} \}
         \eqno(B1)
         $$
         for any $p \in \tilde{\Gamma}$ (see e.g. [2]). Let $p \in
         \Gamma_1 \subset \tilde{\Gamma}$, and suppose that $X \in
         T_p\Gamma_1$ such that
         $$
         \Omega_1 (X,Y) = 0, \hspace{3mm} \forall \hspace{3mm} X \in
         T_p \Gamma_1. \eqno(B2)
         $$
         We will show that $X = 0$. Indeed, let $W \in T_p
         \tilde{\Gamma}$. Using eq. (11), we find a unique vector $U
         \in T_p \Gamma_1$ and $V \in X_p$ such that
         $$
         W = U + V;
         $$
         then
         $$
         \Omega (X,W) = \Omega(X,U) + \Omega (X,V).
         $$
         However, the first term of the R.H.S. vanishes because of
         assumption (B2), the second because of (B1), so
         $\Omega(X,W)
         = 0$ for any $W \in T_p \tilde{\Gamma}$. From (B.1), it
         follows then that $X\in X_p$, so that $X \in X_p \cap T_p
         \Gamma_1$. The transversality (11) implies, however, that
         $X=0$.

         b) We show eq. (12) at a point $p \in \Gamma_1 \subset
         \tilde{\Gamma}$ first. Let $X,Y \in T_p \tilde{\Gamma}$.
         Using eq. (11), we can write
         $$
         X= X_\perp + X_\parallel \hspace{2mm}, \hspace{2mm} Y =
         Y_\perp + Y_\parallel,
         $$
         where
         $$
         X_\perp \in T_p \Gamma_1, Y_\perp \in T_p \Gamma_1,
         X_\parallel \in X_p, Y_\parallel \in X_p
         $$
         and
         $$
         \pi_{1^\star} X = X_\perp \hspace{2mm}, \hspace{2mm}
         \pi_{1^\star} Y = Y_\perp.
         $$
         Then, because of eq. (B1),
         $$
         \tilde{\Omega} (X,Y) = \tilde{\Omega}(X_\perp, Y_\perp) =
         \tilde{\Omega}(\pi_{1^\star} X, \pi_{1^\star} Y) = \Omega_1
         (\pi_{1^\star}X, \pi_{1\star} Y) = (\pi^\star_1 \Omega_1)
         (X,Y).
         $$
         Thus, at $p\in \Gamma_1, \tilde{\Omega} = \pi^\star_1 \Omega_1$.

         Let now $q \in \mbox{D}(\Gamma_1), q \not\in \Gamma_1$.
         Then,
         there is $p\in \Gamma_1$ such that $\pi_1 q= p$, that is, $q$
         and $p$ lie at the same orbits $\gamma_q$.

         As $\gamma_q$ is an integral manifold of the Hamiltonian
         vector fields of the constraints, there will be a symplectic
         diffeomorphism $\varphi$ generated by these vector fields
         that preserves $\gamma_q$, and maps $p$ on $q$. Hence,
         $\varphi$ must satisfy the identity
         $$
         \pi_1 = \pi_1 \circ \varphi,
         $$
         or
         $$
         \pi_{1^\star} = \pi_{1^\star} \circ \varphi_\star.
         $$
         Let $X,Y \in T_q \tilde{\Gamma}$; define
         $$
         U = \varphi^{-1}_\star X, V = \varphi^{-1}_\star Y.
         $$
         Then
         $$
         \tilde{\Omega} (X,Y) = \tilde{\Omega} (\varphi_\star U,
         \varphi_\star V) = \varphi^\star \tilde{\Omega} (U,V) =
         \tilde{\Omega} (U,V).
         $$
         However, $U$ and $V$ are in $T_p\tilde{\Gamma}$, hence
         $$
         \tilde{\Omega}(U,V) = \Omega_1 (\pi_{1^\star} U,
         \pi_{1^\star} V)  = \Omega_1 (\pi_{1^\star}
         (\varphi_{\star}^{-1}U),
         \pi_{1^\star} (\varphi_{\star}^{-1} V)) = \Omega_1(\pi_{1^\star} X,
         \pi_{\star 1}Y)
         $$
         $$
         = (\pi_1^\star \Omega_1) (X,Y),
         $$
         QED.

         \subsection*{B3. Proof of Theorem 4}

         First, we show the following lemma:

         {\bf Lemma}: Let $o$ be a perennial and $X$ its Hamiltonian
         vector field on $(\Gamma, \Omega)$. Let $o_1$ be the
         projection of $o$ to $\Gamma_1$ and $X_1$ the
         Hamiltonian vector field of $o_1$ on $(\Gamma_1, \Omega)_1$.
         Let $p\in \Gamma_1$; let
         $$
         X(p) = X_\perp(p) + X_\parallel(p), \eqno(B3)
         $$
         where $X_\perp (p) \in T_p \Gamma_1$, $X_\parallel(p) = X_p$
         is the unique decomposition of $X(p)$. Then,
         $$
         X_\perp (p) = X_1(p). \eqno(B4)
         $$
         {\em Proof of the Lemma}. The Hamiltonian vector fields $X$
         and $X_1$ at $p$ are defined by the equations
         \[
         \langle \mbox{d} o,Y\rangle = \Omega (Y,X),
         \forall Y \in T_p \Gamma,
         \]
         \[
         \langle \mbox{d} o_1 ,Y_1 \rangle = \Omega_1 (Y_1 ,X_1 ),
         \forall Y_1 \in T_p \Gamma_1 .
         \]
         As $o$ is a perennial, we have $X\in T_p \tilde{\Gamma}$
         and eq. (B3) holds. Moreover,
         \[
         o_1 = i_1^\star \sigma ,
         \]
         $$
         \Omega_1 = i_1^\star \Omega , \eqno(B5)
         $$
         hence
         $$
         \mbox{d} o_1 = i_1^\star \mbox{d} o .
         $$
         Thus, for all $Y_1 \in T_p \Gamma_1$, we obtain that
         \[
         \langle \mbox{d}o_1 ,Y_1 \rangle = \langle \mbox{d}o,
         i_{1\star}Y_1 \rangle = \Omega(i_{1\star}Y_1 ,X) =
         \]
         \[
         \Omega(i_{1\star}Y_1 ,X_{\bot}) = \Omega(i_{1\star}Y_1 ,
         i_{1\star}X_{\bot}) = \Omega_1(Y_1 ,X_{\bot}),
         \]
         QED.

         Returning to the proof of Theorem 4, let us choose two
         arbitrary perennials $o$ and $o'$. Then, clearly
         $$
         i^\star_1 (o + o') = i^\star_1 o + i^\star_1 o',
         $$
         and
         $$
         i^\star_1 (o\cdot o') = (o\cdot o')\mid_{\Gamma_1} = o
         \mid_{\Gamma_1} \cdot o'\mid_{\Gamma_2} = (i^\star_1 o) \cdot
         (i^\star_1 o').
         $$
         Finally,
         $$
         \{ o_1, o'_1 \}_1 \mid_p = \Omega_1 (X_1, X'_1) \mid_p = \Omega
         (X_1, X_1') \mid_p.
         $$
         Here, we have used eq. (B5) and
         that $X_1$ and $X_1'$ are tangential to $\Gamma_1$. Then,
         using the Lemma and the relation (B1) we obtain:
         $$
         \Omega(X_1, X'_1) \mid_p = \Omega (X_\perp, X_\perp') \mid_p =
         \Omega (X_\perp + X_\parallel, X_\perp' +X'_\parallel) \mid_ p =
         $$
         $$
         = \Omega (X,X')\mid_p = \{ o, o'\}\mid_p.
         $$
         Hence: $i^\star_1$ preserves alls operations of the Poisson
         algebra. The claim 1) of Theorem 4 follows immediately, as
         well as the claim that $i^\star_1$ is a Poisson algebra
         homomorphism.

         Finally, any perennial $o$ such that $i^\star_1o$ is
         zero everywhere on $\mbox{D}(\Gamma_1)$ means that $o$
         itself vanishes on $\mbox{D}(\Gamma_1)$, QED.

         \subsection*{B4. Proof of Theorem 6}

         The group $\varphi_{1t}$ is related to $\varphi_t$ by
         $$
         \varphi_{1t} = \pi_1 \circ \varphi_t \circ i_1.
         $$
         From this, the relation between the generators $X_1$ and $X$
         follows:
         $$
         X_1 = \pi_{1^\star}X.
         $$
         Hence, $X_1 = X_\perp$ and
         $$
         X = X_\parallel + X_1.
         $$
         Then, the Lemma of subsection B3 implies that $X_1$
         is the Hamiltonian vector field of $f_1$ in $(\Gamma_1,
         \Omega_1)$,
         QED.

         \subsection*{B5. Proof of Theorem 7}

         We can write $\sigma$ as follows
         $$
         \sigma = \pi_2 \circ i_1 \eqno(B6)
         $$
         and $\sigma^{-1}$ as follows
         $$
         \sigma^{-1} = \pi_1 \circ i_2. \eqno(B7)
         $$
         Then, using eq. (12),
         $$
         \sigma^\star \Omega_2 = i^\star_1 (\pi^\star_2 (\Omega_2)) =
         i^\star_1 \tilde{\Omega} = \Omega_1,
         $$
         and, using the property of perennials that
         $$
         \pi^\star_2 (o_2) = o_2 \mid_{\tilde{\Gamma}},
         $$
         we have
         $$
         \sigma^\star o_2 = i^\star_1 (\pi^\star_2 (o_2)) = i^\star
         _1 (o) = o_1.
         $$
         To show the point c), we substitute for $\sigma$ and
         $\sigma^{-1}$ from (B6) and (B7):
         $$
         \sigma \circ \varphi_1 \circ \sigma^{-1} = \pi_2 \circ i_1
         \circ \varphi_1 \circ \pi_1 \circ i_2 = \pi_2 \circ \varphi_1
         \circ
         \pi_1 \circ i_2,
         $$
         as $i_1 \circ \varphi_1 = \varphi_1$. If we use the definition
         of $\varphi_1$,
         $$
         \varphi_1 = \pi_1 \circ \varphi \circ i_1,
         $$
         we obtain, as $i_1 \circ \pi_1 = \pi_1$, that
         $$
         \sigma \circ \varphi_1 \circ \sigma^{-1} = \pi_2 \circ \pi_1
         \circ \varphi \circ \pi_1 \circ i_2.
         $$
         However,
         $$
         \pi_1 \circ \varphi \circ \pi_1 = \pi_1 \circ \varphi,
         $$
         as $\pi_1$ moves points along orbits and $\varphi$ maps
         orbits onto orbits. Similarly,
         $$
         \pi_2 \circ \pi_1 = \pi_2,
         $$
         hence
         $$
         \sigma \circ \varphi_1 \circ \sigma^{-1} = \pi_2 \circ
         \varphi \circ i_2 = \varphi_2,
         $$
         QED.
         \subsection*{B6. Proof of Theorem 9}

         A) Consider the curve $\xi_\gamma(t)$ in $\Gamma_0$. Let
         $\gamma_t$ be the orbit that contains the point
         $\xi_\gamma(t)$. Then, as $h(t) \xi_\gamma(t) \in
         \gamma$, and $h(t)$ maps orbits onto orbits, we have
         $$
         \gamma_t = h^{-1}(t) \gamma.
         $$
         Next consider two points $p = \xi_\gamma(t)$ and
         $\xi_\gamma (t + \Delta t)$; as
         $$
         \gamma_{t+\Delta t} = h^{-1}(t+\Delta t) \gamma.
         $$
         we obtain that
         $$
         \gamma_{t+ \Delta t} = h^{-1} (\Delta t) \gamma_t.
         $$
         Hence, the two points $\xi_\gamma (t + \Delta_t)$ and
         $h^{-1} (\Delta t)p$ lie at the same orbit
         $\gamma_{t+\Delta t}$. Going to the limit $\Delta t
         \rightarrow 0$, we obtain the relation:
         $$
         \lim_{\Delta t \rightarrow 0} \frac{h^{-1}(\Delta t)p}
         {\Delta t} =
         \frac{d \xi_\gamma(t)}{dt} + X_\parallel,
         $$
         where $X_\parallel$ is a vector tangential to $\gamma_t$. The
         vector on L.H.S is, however $- X(p)$, $X$ being the Hamiltonian
         vector field of $h$, so the equation above is equivalent to
         $$
         (-X(p))_\perp = \frac{d \xi_\gamma(t)}{dt}.
         $$
         Then, the Lemma of Sec. B3 implies
         $$
         \frac{d \xi_\gamma(t)}{dt} = - X_0(p)
         $$
         where $X_0$ is the Hamiltonian vector field of the
         projection, $-H$, of $h$ to $(\Gamma_0, \Omega_0)$, and the
         A-part of Theorem 9 is shown.

         B) Eq. (13) implies that
         $$
         o_t (h(t)p) = o(p), \hspace{2mm} p\in \Gamma_0.
         $$
         Let $\gamma$ be the orbit through $p$. Then, as $o_t$ is a
         perennial, and as $\xi_\gamma(-t)$ lies at the same orbit
         as $h(t)p$, we obtain
         $$
         \tilde{o}_t (\xi_\gamma (-t)) = o_t (h(t)p) = \tilde{o}(p).
         $$
         The A-part of Theorem 9 implies then
         $$
         \frac{d\tilde{o}_t}{dt}\mid_{\xi_\gamma(-t)} - \{
         \tilde{o}_t, H \}\mid_{\xi_\gamma (-t)} = 0.
         $$
         This implies eq. (15) immediately, QED.

         \end{document}